\documentclass[preprintnumbers,article,amsmath,amssymb,floatfix,10pt,prd,superscriptaddress,nofootinbib,showkeys]{revtex4}

\usepackage{doi}
\usepackage{hyperref}
\hypersetup{
	colorlinks=true,        
	linkcolor=blue,         
	citecolor=cyan,         
}

\usepackage{graphicx}
\usepackage{dcolumn}
\usepackage{bm}
\usepackage{color}
\usepackage{enumitem}
\usepackage{amsmath}
\usepackage{amssymb}

\newcommand{\beq}{\begin{equation}}
\newcommand{\eeq}{\end{equation}}
\newcommand{\bea}{\begin{eqnarray}}
\newcommand{\eea}{\end{eqnarray}}

\usepackage{bbm}
\usepackage{amsfonts}
\usepackage{mathrsfs}
\usepackage{latexsym}
\usepackage{epsfig}
\usepackage{epstopdf}
\usepackage{epstopdf}
\usepackage{graphicx}
\usepackage{amssymb}
\usepackage{amsmath}
\usepackage{dcolumn}
\usepackage{bm}
\usepackage{color}
\usepackage{comment}
\usepackage{float}
\usepackage{xcolor}
\usepackage{orcidlink}

\begin{document}

\preprint{}

\title{Yukawa-Casimir wormholes within Einstein-Cartan gravity framework}

\author{Nayan Sarkar
\orcidlink{0000-0002-3489-6509}}
\email{ nayan.mathju@gmail.com}
\affiliation{Department of Mathematics, Karimpur Pannadevi College, Nadia 741152, West Bengal, India}

\author{Susmita Sarkar\orcidlink{0009-0007-1179-2495}}
\email{ susmita.mathju@gmail.com}
\affiliation{Department of Applied Science and Humanities, Haldia Institute of Technology, Haldia-721606, West Bengal, India}

\author{Abdelmalek Bouzenada\orcidlink{0000-0002-3363-980X}}
\email{ abdelmalekbouzenada@gmail.com}
\affiliation{Laboratory of Theoretical and Applied Physics, Echahid Cheikh Larbi Tebessi University 12001, Algeria}

\date{\today}

\begin{abstract}
The Einstein-Cartan (EC) theory of gravity provides a natural extension of general relativity by incorporating spacetime torsion to account for the intrinsic spin of matter. In this work, we investigate Yukawa-Casimir traversable wormholes supported by three distinct Yukawa-Casimir energy density profiles within the framework of EC gravity. The resulting shape functions are shown to satisfy all the fundamental requirements for traversable wormhole geometries. Our analysis reveals that the presence of exotic matter is unavoidable in sustaining these wormholes, and we quantify its total amount through the volume integral quantifier. Furthermore, the equilibrium of the wormhole configurations is established by examining the Tolman-Oppenheimer-Volkoff equation. To enhance the physical relevance of the present work, we study several key features of the wormholes, including the embedding surface, proper radial distance, tidal forces, and total gravitational energy. In addition, we analyze the optical properties of wormholes by examining both the shadow and the strong deflection angle. All the findings collectively demonstrate the physical plausibility of Yukawa-Casimir traversable wormholes within the EC gravity framework.
\end{abstract}


\maketitle

\section{Introduction}\label{sec1}

Wormholes are widely conceived in theoretical physics as hypothetical tunnel-like configurations of spacetime that can act either as conduits connecting two otherwise distant regions of the same universe or as bridges that establish communication between entirely distinct universes, and the modern theoretical framework for traversable wormholes was firmly established in the landmark work of Morris and Thorne in 1988 \cite{WHM1, WHM2}, where they introduced a class of solutions that are traversable by observers under suitable conditions, the most fundamental requirement being the so-called flaring–out condition at the wormhole throat, which ensures the stability and openness of the geometry but simultaneously implies a violation of the null energy condition (NEC) when considered within the standard framework of Einstein’s GR, matter fields that exhibit such a violation are collectively described as exotic matter \cite{WHM3, WHM4}, and the unavoidable presence of such matter constitutes one of the most serious challenges in the construction of realistic wormhole scenarios, since it conflicts with the standard energy conditions that play a pivotal role in classical gravitational physics, which has motivated a large number of researchers to explore strategies aimed at minimizing the use of exotic matter \cite{WHM5}, among which one of the most influential approaches is the thin-shell formalism developed by Visser and Poisson, in which the exotic matter is confined strictly to the wormhole throat, thereby reducing its role in the surrounding spacetime \cite{WHM6}, and since then thin-shell wormholes have been subjected to extensive study regarding their stability, physical interpretation, and applications across a wide range of theoretical models \cite{WHM7, WHM8, WHM9, WHM10, WHM11, WHM12, WHM13, WHM14, WHM15, WHM16}, alongside this line of research, significant attention has also been devoted to the construction of wormhole geometries in alternative and extended theories of gravity, with the aim of alleviating or eliminating the necessity of exotic matter, and numerous important results have been reported in this context, including traversable wormhole solutions in Brans–Dicke theory \cite{WHM17, WHM18, WHM19, WHM20}, wormhole models formulated in the framework of $f(R)$ gravity \cite{WHM21, WHM22, WHM23}, solutions inspired by Born–Infeld theory \cite{WHM24,WHM25}, higher-order Lovelock gravity \cite{WHM26}, as well as in higher-dimensional or unified field approaches such as Kaluza–Klein ($5\mathcal{D}$) gravity \cite{WHM27, WHM28} and scalar–tensor theories \cite{WHM29}, while further studies have demonstrated that wormhole solutions may also be realized in spacetimes endowed with a cosmological constant \cite{WHM30, WHM31}, in more recent years, the exploration of wormholes in $f(R,T)$ gravity, where the gravitational action depends simultaneously on the Ricci scalar $R$ and the trace of the energy–momentum tensor $T$, has gained increasing attention, leading to the discovery of diverse classes of wormhole solutions including those supported by different types of Chaplygin gas \cite{WHM32}, configurations that satisfy the conventional energy conditions within exponential $f(R,T)$ gravity models \cite{WHM33}, as well as a variety of other extensions investigated in detail in the literature \cite{WHM34, WHM35, WHM36, WHM37, WHM38}, and it has been shown that such wormhole geometries may persist over either extremely short or arbitrarily long temporal intervals depending on the chosen conditions \cite{WHM39,WHM40}, while in certain cases they can even satisfy the dominant energy condition throughout the entire spacetime manifold \cite{WHM41}, an achievement that directly challenges the conventional association of wormholes with exotic matter, Also, this research direction has been extended into the domain of non-static or dynamical wormholes, where the throat evolves with time and the geometrical structure must be analyzed in a dynamical framework, and these models have provided evidence that wormhole configurations may satisfy not only the pointwise but also the averaged forms of the energy conditions when evaluated along timelike or null geodesics over specific time intervals \cite{WHM42, WHM43}, thereby opening a promising avenue toward reconciling wormhole physics with realistic physical properties and illustrate the potential role of modified gravity theories and novel matter sources influence in spacetime structures.

The Casimir effect (CE) is a significant illustration of quantum vacuum fluctuations, demonstrating that space is characterized by dynamic fields rather than being truly empty. It was originally proposed by Casimir in 1948 \cite{CE1} and in another application confirmed by a sequence of experimental studies, ranging from the early measurements by Sparnaay \cite{CE2}, to the precision experiments of Mohideen and Roy \cite{CE3}, the work of Bressi and collaborators \cite{CE4}, and more modern investigations focusing on quantum vacuum properties \cite{CE5, CE6}. Fundamentally, the effect occurs when two parallel, neutral, conducting plates placed in close separation inside a vacuum alter the allowed electromagnetic field modes, producing an imbalance in the zero-point energy of the quantum electrodynamics (QED) vacuum and resulting in an attractive force between the plates. This macroscopic quantum phenomenon, frequently interpreted as a manifestation of negative energy density, has motivated numerous investigations into its possible role in wormhole physics. In particular, Garattini studied the possibility of sustaining Morris–Thorne-type wormholes solely with Casimir energy while examining the implications of the Quantum Weak Energy Condition (QWEC) on their traversability \cite{CE7}. In this context, further work by Garattini \cite{CE8, CE9} introduced the notion of Yukawa–Casimir wormholes, where Yukawa-type corrections are incorporated into the Casimir source, aiming to provide exotic energy distributions capable of producing zero-tidal-force wormholes. Also, this line of investigation was later broadened by Jawad et al. \cite{CE10} and Mishra et al. \cite{CE11}, who examined the stability and physical feasibility of such wormhole solutions. Another important result in the context of these developments is that the present study \cite{CE11} illustrates wormhole configurations supported by modified Casimir energy densities with Yukawa-type corrections, deepening the connection between quantum vacuum phenomena and traversable wormhole geometries. 

In 1923, E. Cartan proposed a fundamental modification of Einstein’s GR, which is today known as the Einstein-Cartan (EC) theory \cite{CG1, CG2}, a theoretical framework that introduced a direct relationship between the intrinsic angular momentum of matter and the torsion of spacetime, anticipating by two years the concept of spin in quantum theory formally established by Goudsmit et al. in 1925. Within the general relativistic setting, there are essentially two different ways to incorporate spin into the theory. In this case, the first approach consists of treating spin as an additional dynamical variable while keeping the background geometry purely Riemannian, thereby preserving the standard curvature-based structure of spacetime \cite{CG3, CG4, CG5, CG6, CG7}, in this case, the spin resembles that of quantum mechanics and naturally corresponds to the spin described in the Dirac theory of the electron, offering a framework in which spinning particles may be analyzed in curved backgrounds without altering the geometric foundation of GR. The second, and more profound, approach is that developed by Cartan, in which the structure of spacetime itself is generalized so that it possesses both curvature and torsion through the assumption that the metric tensor and the affine connection are independent objects \cite{CG8}. Also, the resulting geometric model, which is Riemann-Cartan geometry or $U_4$, differs from ordinary Riemannian geometry by including torsion as an intrinsic property of the manifold. Building on Cartan’s seminal insights, many physicists extended and clarified this formalism. Hehl and collaborators developed a systematic treatment of torsion and its coupling with spin \cite{CG9, CG10, CG11, CG12}. Trautman contributed to the mathematical structure of the theory \cite{CG13, CG14}, and Kopczynski further investigated the dynamical role of spin-torsion interactions \cite{CG15, CG16}. A key feature of EC theory is that torsion is not dynamical in the sense of propagating degrees of freedom,  rather, it is completely illustrated algebraically by the spin distribution of matter, vanishing in the absence of spin and thereby representing a local effect sourced exclusively by particles with intrinsic angular momentum \cite{CG9, CG10, CG11, CG12}. In cosmological contexts, the inclusion of torsion and spin has been widely studied through the model of the so-called Weyssenhoff fluid, which represents a natural generalization of the perfect fluid to include spin effects. In this context, this semiclassical fluid is characterized not only by its energy density $\rho_s$ and pressure $p_s$, but also by a spin density vector $S^\alpha$ that is constrained to remain orthogonal to the four-velocity $u^\alpha$ of the fluid in the comoving reference frame, thereby ensuring consistency with relativistic kinematics. Another important result in this context, introduced and developed by Weyssenhoff and collaborators, provided a tractable model for analyzing the dynamics of matter endowed with spin in curved spacetime \cite{CG17, CG18}. In this context, its incorporation into the Einstein-Cartan framework has revealed significant physical information, particularly in the early universe description, where high densities and strong gravitational fields show the effects of torsion, potentially alleviating singularities, modifying cosmological evolution, or influencing the behavior of quantum fields in curved backgrounds \cite{CG19, CG20}.  

The analysis of energy conditions (EC) constitutes a fundamental aspect in exploring the viability of alternative formulations of gravitation beyond Einstein’s General Relativity, since they act as diagnostic tools to distinguish physically acceptable models from those leading to exotic or non-standard behaviors. Over the last decades, a wide range of contributions has enriched this field by examining how the weak, strong, null, and dominant EC manifest in different modified gravity scenarios. For instance, Capozziello et al. \cite{EC1} provided a comprehensive study of $f(R)$ gravity under a power-law prescription of the Ricci scalar, illustrating how such models influence the standard energy bounds. In a related direction, Atazadeh et al. \cite{EC2} investigated the role of EC in the Brans-Dicke scalar-tensor framework, which can be derived from a generic $f(R)$ model, and showed that constraints on the scalar field dynamics emerge naturally from these considerations. Attention has also been devoted to torsional modifications of gravity, in particular, Liu and Rebouças \cite{EC3} examined $f(T)$ gravity, including exponential and Born-Infeld inspired forms, and analyzed how the imposition of EC impacts cosmological viability. Another important result: Zubair et al. \cite{EC4} discussed a theory where matter couples non-minimally to the torsion scalar and reported that such a coupling significantly alters the structure of traditional energy bounds. In this context, Azizi et al. \cite{EC5} tested higher-derivative ($\mathcal{D}$) torsion gravity, emphasizing that extended derivative terms strongly affect the validity of energy inequalities. Also, curvature invariants beyond the Ricci scalar have been widely studied, Garcia et al. \cite{EC6} and Bamba et al. \cite{EC7} both considered $f(G)$ gravity with $G$ representing the Gauss-Bonnet invariant, and they showed how different forms of $f(G)$ can either preserve or violate the classical EC, thereby affecting late-time acceleration models. In continuation of this line, Sharif and Ikram \cite{EC8} analyzed reconstructed $f(G, T)$ models in a FLRW cosmological background, clarifying how the mixture of geometry and matter terms influences the fulfilment of energy criteria. Yousaf et al. \cite{EC9} focused on higher-order gravity theories such as $f(R, \square R, T)$ gravity, demonstrating that the imposition of EC restricts the parameter space and determines the physical acceptability of the theory. More recently, attention has shifted toward symmetric teleparallel gravity, where Mandal et al. \cite{EC10} testing $f(Q)$ gravity with $Q$ illustrate the non-metricity scalar, showing that EC provides essential guidelines for ensuring observational and theoretical consistency. This framework was further generalized by Arora et al. \cite{EC11} in the context of $f(Q, T)$ gravity, where they established that incorporating matter contributions explicitly alongside the non-metricity scalar results in novel modifications of energy inequalities that shape cosmic evolution.

Our study is dedicated to exploring Yukawa-Casimir traversable wormholes within the framework of EC gravity. The paper is structured as follows: Section~\ref{sec1} provides a general introduction, while Section~\ref{sec2} outlines the formalism of EC gravity. In Section~\ref{sec3}, we derive the wormhole field equations in this framework, and Section~\ref{sec4} is devoted to the discussion of the energy conditions. The Yukawa-Casimir energy density profiles are presented in Section~\ref{sec5}, followed by Section~\ref{sec6}, where we construct wormhole solutions supported by three such profiles. The continuity between the interior wormhole geometry and the exterior Schwarzschild spacetime is examined in Section~\ref{sec7}. Section~\ref{sec8} investigates the total amount of exotic matter through the volume integral quantifier ($\mathcal{VIQ}$), while Section~\ref{sec9} focuses on the equilibrium analysis of the obtained wormhole configurations. In Section~\ref{sec10}, we explore several physical characteristics of the wormholes, including the embedding surface and proper radial distance in Subsection~\ref {sec10a}, the tidal forces in Subsection~\ref {sec10b}, and the total gravitational energy in Subsection~\ref {sec10c}. In this case, Section-\ref{sec11} and Section-\ref{sec12} tested the shadow and the strong deflection angle of the wormholes, respectively. Finally, Section \ref{sec13} summarizes the key findings and conclusions of this work..

\section{Basic equations in Einstein-Cartan gravity}\label{sec2}
The Einstein-Cartan (EC) theory of gravity is regarded as the most direct and fundamental generalization of Einstein’s general relativity, where the intrinsic spin of matter is naturally accommodated through spacetime torsion. Within this formulation, the action integral takes the following form
\begin{eqnarray}
S &=&\int d^4x \sqrt{g}\left[\frac{-1}{2k}(R+2\Lambda)+\mathcal{L}_m\right]
=\int d^4x \sqrt{g}\Big[\frac{-1}{2k}\left\{R(\{\})+C^\alpha_{\beta\lambda}C^{\beta\lambda}_\alpha-C^\alpha_{\beta\alpha}C^{\beta\lambda}_\lambda+2\Lambda\right\}+\mathcal{L}_m\Big],\label{s}
\end{eqnarray}
where $k = 8\pi G/c^4$ denotes the gravitational coupling constant,  $R$ is the Ricci scalar, $\mathcal{L}_m$ represents the Lagrangian of the matter fields, and $\Lambda$ corresponds to the cosmological constant, having the dimension of $length^{-2}$. Moreover, the tensor $C^\mu_{\alpha\beta}$, known as the contortion tensor, is defined as
\begin{equation}
C^\mu_{\alpha\beta}=T^\mu_{\alpha\beta}+T_{\alpha\beta}^{~~\mu}+T_{\beta\alpha}^{~~\mu}.\label{C}
\end{equation}
Here, $T^\alpha _{\mu\nu}$ represents the spacetime torsion tensor, which is defined  as the anti-symmetric part of the connection
\begin{equation}
T^\mu_{\alpha\beta}=\frac{1}{2}\left[\Gamma^\mu_{\alpha\beta}-\Gamma^\mu_{\beta\alpha}\right].\label{C}
\end{equation}

The Cartan field equation is derived by varying the action (\ref{s}) with respect to the contortion tensor, yielding
\begin{equation}
T^\alpha_{\mu\beta}-\delta^\alpha_\beta T^\gamma_{\mu\gamma}+\delta^\alpha_\mu T^\gamma_{\beta\gamma}=-\frac{1}{2} k \tau_{\mu\beta}^{~~\alpha}, \label{t1}
\end{equation}
where $\tau^{\mu\alpha\beta} = 2(\delta\mathcal{L}_m/\delta C_{\mu\alpha\beta})/\sqrt{-g}$  represents the spin tensor of matter \cite{fw76}. It is important to note that, since the field equation for the torsion tensor is algebraic, it does not permit the generation of torsion waves outside the matter distribution \cite{fw76}. Consequently, spacetime torsion is restricted to the interior of matter configurations. Furthermore, variation of the action (\ref{s}) with respect to the metric leads to the EC field equation, expressed as \cite{fw76, vd86, vd90, vd94}
\begin{equation}
G_{\mu\nu}(\{\}) - \Lambda g_{\mu\nu}=\kappa(T_{\mu\beta}+\theta_{\mu\beta}),\label{G}
\end{equation}
where $()$ denotes symmetrization and $\theta_{\mu\beta}$ can be defined in the following form
\begin{eqnarray}
    \theta_{\mu\beta}=\frac{1}{\kappa}\Big[4T^\eta_{~\mu\eta}T^\beta_{~\mu\beta}-\left(T^\rho_{~\mu\epsilon}+2T_{(\mu\epsilon)}^{~~~\rho}\right)\left(T^\epsilon_{~\nu\rho}+
    2T_{(\nu\rho)}^{~~~\epsilon}\right)+\frac{g_{\mu\nu}}{2}\left(T^{\rho\sigma\epsilon}+2T^{(\sigma\epsilon)\rho}\right)\left(T_{\epsilon\sigma\rho}+2T_{(\sigma\rho)\epsilon}\right)-2g_{\mu\nu}T^{\rho\sigma}_{~\rho}T^{\sigma}_{~\epsilon\sigma}\Big].\label{theta}
\end{eqnarray}
 Here, the tensor $\theta_{\mu\nu}$ accounts for the correction to the dynamical energy-momentum tensor $T_{\mu\beta} = 2(\delta \mathcal{L}_m/\delta g^{\mu\beta})/\sqrt{-g}$ arising from the spin contributions to the spacetime geometry. It should be noted that if the matter fields do not depend on spacetime torsion, i.e., when $\theta_{\mu\nu}$ = 0, the field equation (\ref{G}) reduces to the standard Einstein field equation with a cosmological constant. However, the field equation (\ref{G}) can also be expressed as
\begin{eqnarray}
    R_{\mu\nu}-\frac{1}{2}Rg_{\mu\nu}-\Lambda g_{\mu\nu} = \kappa\Delta_{\mu\nu}, \label{eq}
\end{eqnarray}

 where $R_{\mu\nu}$ is the Ricci scalar, and $\Delta_{\mu\nu}$ denotes the canonical energy-momentum tensor, which is related to $T_{\mu\beta}$  through the Belinfante-Rosenfeld relation  as
\begin{eqnarray}
    \Delta_{\alpha\beta}=T_{\alpha\beta}+\frac{1}{2}\left(\nabla_\mu-2T^\gamma_{~\mu\gamma}\right)\left(\tau_{\alpha\beta}^{~\mu}-\tau_{\beta~\alpha}^{~\mu}+\tau^\mu_{~~\alpha\beta}\right).\label{delta}
\end{eqnarray}

where $\nabla_\mu$  represents the covariant derivative concerning the asymmetric connection \cite{nj34}. It is worth noting that the Bianchi identities (\ref{t1}), together with the EC field equations (\ref{eq}), simultaneously yield the conservation laws for both the canonical energy–momentum tensor and the spin tensor \cite{fw76, tw61, dw62, dw64, dw64a, fw71, fw71a, rt02, dn16, ea76, fw73, fw74, yn87}.

\begin{figure}[h]
\begin{center}
\begin{tabular}{rl}
\includegraphics[width=5.6cm]{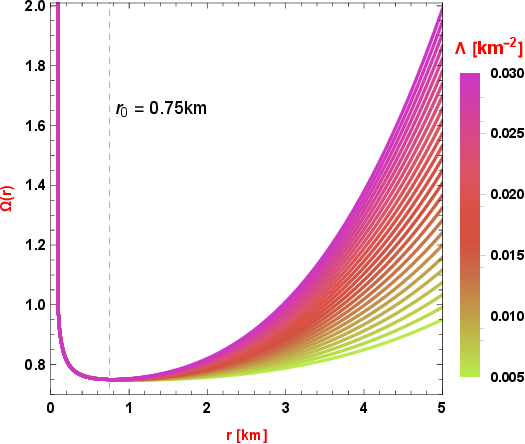}
\includegraphics[width=5.6cm]{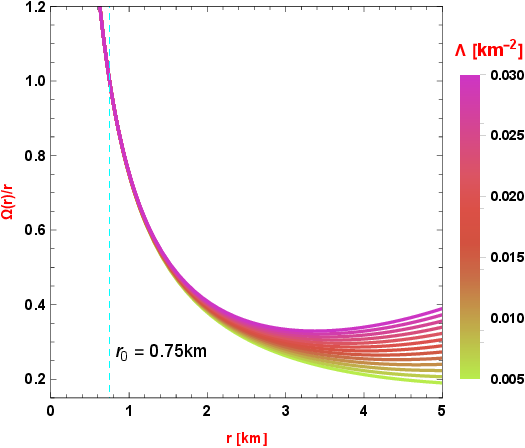}
\includegraphics[width=5.6cm]{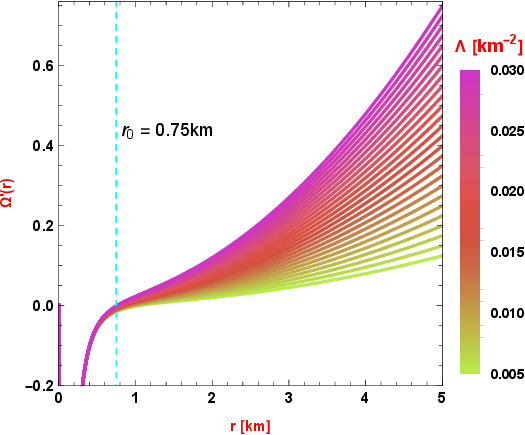}
\\
\end{tabular}
\end{center}
\caption{\label{fig1} Depiction of the shape function $\Omega(r)$ (Left), $\Omega(r)/r$ (Middle), $\Omega'(r)$ (Right) against the radial coordinate $r$ for the energy density model-I with parameters $r_0 = 0.75 km$, $\alpha = -0.8 km$, $S_0 = 0.01 km^{-1}$, and $\lambda = 1.45km^{-1}$.}
\end{figure}

\section{Einstein field equations in wormhole spacetime with a modified source}\label{sec3}

To investigate wormhole solutions within the framework of EC theory, it is essential to incorporate the role of intrinsic spin into the matter distribution. For this purpose, we adopt the classical description of spin proposed by Weyssenhoff, in which matter is modeled as a fluid endowed with microscopic intrinsic angular momentum. In this representation, the so-called Weyssenhoff fluid behaves like a continuous medium whose spin density couples directly to spacetime torsion. Accordingly, the spin tensor is expressed as \cite{yn87, jw47, jr83, ga74}
\begin{eqnarray}
\tau_{\mu\nu}^{~~\alpha} = S_{\mu\nu} u^\alpha, \quad\quad S_{\mu\nu}u^\mu = 0,\label{S}
 \end{eqnarray}

where $u^\alpha$ represents the four-velocity and $S_{\mu\nu}$ stands for the spin density tensor.  

The total energy-momentum tensor incorporates both the conventional fluid component and an additional term arising from intrinsic spin, expressed as \cite{MG86}
\begin{eqnarray}
    T^{total}_{\alpha\beta} &=& T_{\alpha\beta}+\theta_{\alpha\beta}\nonumber
    \\
    &=&\left[(\rho+P_t)u_\alpha u_\beta+P_tg_{\alpha\beta} + (P_r-P_t)v_\alpha v_\beta\right]+ u_{(\alpha}S_{\beta)}^{~\mu}u^\nu C^\rho_{~\mu\nu}u_\rho+u^\rho C^\mu_{~\rho\sigma}u^\sigma u_{(\alpha}S_{\beta)\mu}-\frac{1}{2}u_{(\alpha}T_{\beta)\mu\nu}S^{\mu\nu}\nonumber
    \\
    &&+\frac{1}{2}T_{\mu\nu(\alpha}S^\mu_{~\beta)}u^\nu,\label{Ttotal}
\end{eqnarray}

where $v_\mu$ denotes a unit spacelike vector field along the radial direction, and $\rho = \rho(r)$ denotes the energy density, $P_r = P_r(r)$ represents the radial pressure, $P_t = P_t(r)$ corresponds to the tangential pressure of the matter distribution. Now, combining Eqs. (\ref{G}), (\ref{theta}), (\ref{S}), and (\ref{Ttotal}) leads to Einstein’s field equation with an anisotropic matter distribution and additional spin–correction terms, given by
\begin{eqnarray}
    G_{\mu\nu}-\Lambda g_{\mu\nu} &=&\kappa \left[\rho(r) + P_t(r)-\frac{\kappa}{2} S^2\right]u_\mu u_\nu+\kappa\Big[P_t(r)-\frac{\kappa}{4}S^2\Big] g_{\mu\nu}+\left[P_r(r)-P_t(r)\right]v_\mu v_\nu,\label{EE}
\end{eqnarray}

where $S^2 = \frac{1}{2} \langle S_{\mu\nu} S^{\mu\nu} \rangle$.  

\begin{figure}[h]
	\begin{center}
		\begin{tabular}{rl}
			\includegraphics[width=6cm]{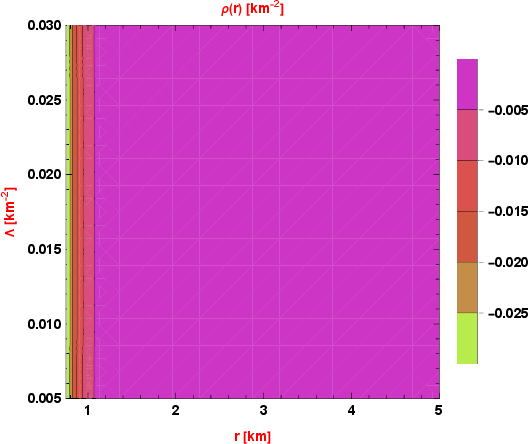}
			\includegraphics[width=5.6cm]{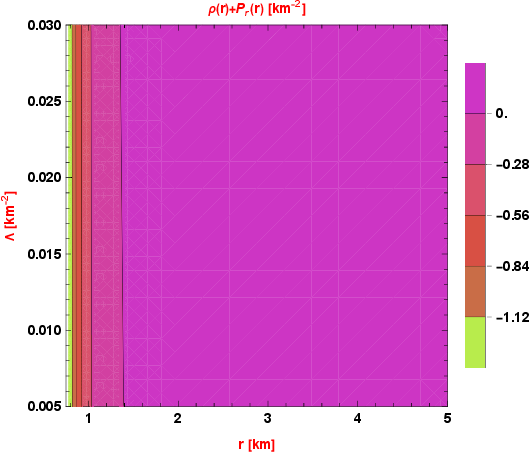}
			\includegraphics[width=5.7cm]{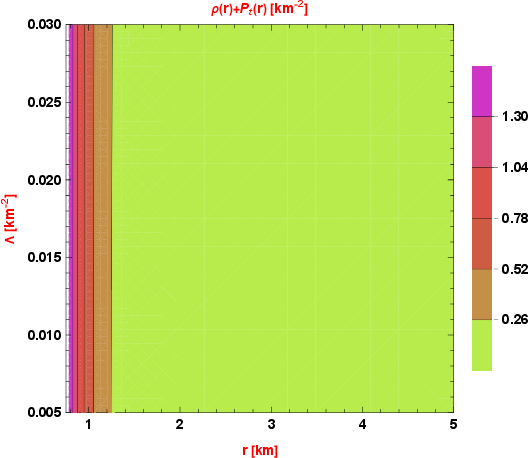}
			\\
		\end{tabular}
		\begin{tabular}{rl}
			\includegraphics[width=5.9cm]{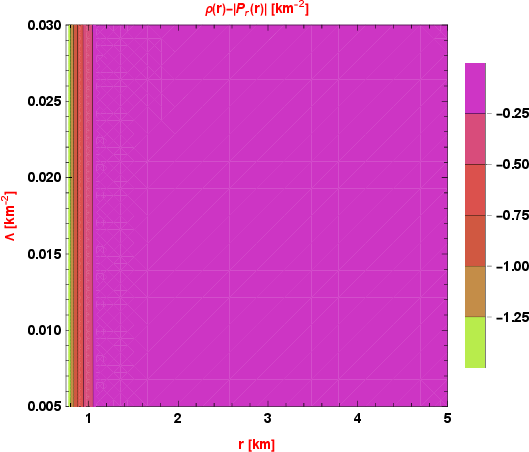}
			\includegraphics[width=5.9cm]{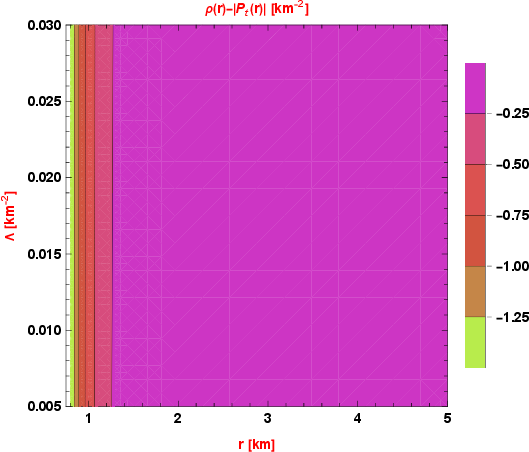}
			\includegraphics[width=5.7cm]{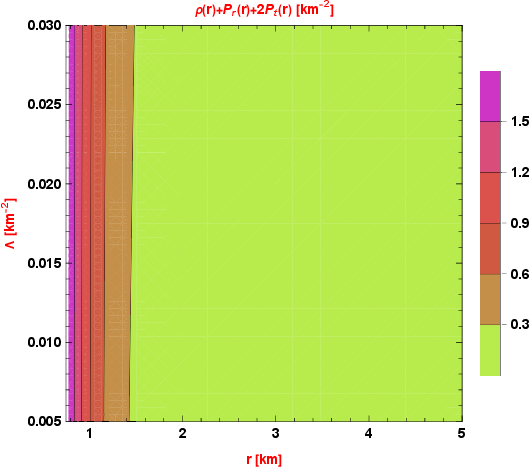}
			\\
		\end{tabular}
	\end{center}
	\caption{ \label{fig2} Depiction of the energy density $\rho(r)$ (Left), $\rho(r)+P_r(r)$ (Middle), $\rho(r)+P_t(r)$ (Right) in the above panel, and  $\rho(r)-|P_r(r)|$ (Left), $\rho(r)-|P_t(r)|$ (Middle), $\rho(r)+P_r(r)+2P_t(r)$ (Right) in the below panel for the energy density model-I with parameters $r_0 = 0.75 km$, $\alpha = -0.8 km$, $S_0 = 0.01 km^{-1}$, and $\lambda = 1.45km^{-1}$.}
\end{figure}

To formulate Einstein’s field equation (\ref{EE}) in the context of a wormhole geometry, we adopt the Morris-Thorne line element for a traversable wormhole, given by \cite{MT88}
\begin{equation}
ds^2 = -e^{2\Phi(r)}dt^2+\left(1-\frac{\Omega(r)}{r}\right)^{-1}dr^2+r^2(d\theta^2 + sin^2\theta d\phi^2),\label{Metric}
\end{equation}

where $\Phi(r)$ and $\Omega(r)$ are referred to as the redshift function and the shape function, respectively. The minimum radius $r = r_{0}$, satisfying the condition $\Omega(r_{0}) = r_{0}$, defines the throat of the wormhole. For a wormhole to be traversable, the following conditions must be fulfilled:  The redshift function $\Phi(r)$ must remain finite everywhere so that no event horizon is present,  the shape function $\Omega(r)$ must satisfy the condition (i) $\frac{\Omega(r)}{r} \leq 1$ for $r \geq r_0$ and (ii) the flare-out condition, defined as $\Omega^{\prime}(r) < 1$ for $r\geq r_0$.

The Einstein field equations (\ref{EE}) corresponding to the wormhole metric (\ref{Metric}) can now be expressed as, under the choice of units $\kappa=c=1$

\begin{eqnarray}
\rho(r) &=& \frac{1}{4r^2}\left[4\Omega^\prime(r)+r^2 S^2(r)-4\Lambda r^2\right], \label{den}\\
 P_r(r) &=& \frac{1}{4r^3}\Big[ 8r\Phi'(r)\left(r-\Omega(r)\right)-4\Omega(r)+r^3 S^2(r)+4\Lambda r^3\Big], \label{pr}\\
 P_t(r) &=& \frac{1}{4r^3}\big[ 4r^2\left(\Phi''(r)+\Phi'^{2}(r)\right)\left(r-\Omega(r)\right)-2r\Phi'(r)\big(r\Omega'(r)-2r+B(r)\big)+r^3S^2(r)-2r\Omega'(r)+4\Lambda r^3+2\Omega(r)\big]\nonumber,\label{pt}
 \\
\end{eqnarray}
where $\prime$ represents the derivative with respect to the radial coordinate $r$. In this framework, the conservation law for the total energy-momentum tensor takes the following form
\begin{eqnarray}
\Phi'(r)\left[\rho(r)+P_r(r)\right]+P_r'(r)+\frac{2}{r}\left[P_r(r)-P_t(r)\right]-\frac{1}{2}\left[\Phi'(r)S^2+\frac{1}{2}(S^2)'\right]=0.\label{cq}
\end{eqnarray}

\begin{figure}[h]
	\begin{center}
		\begin{tabular}{rl}
			\includegraphics[width=5.6cm]{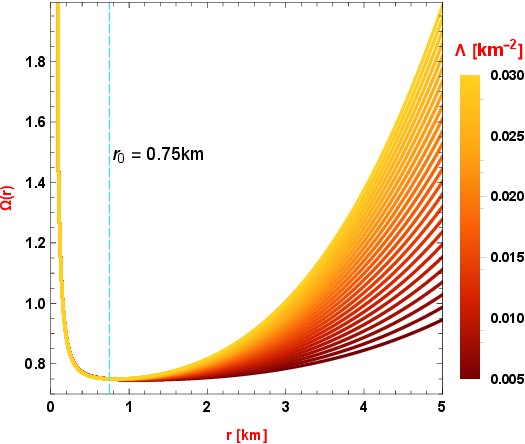}
			\includegraphics[width=5.6cm]{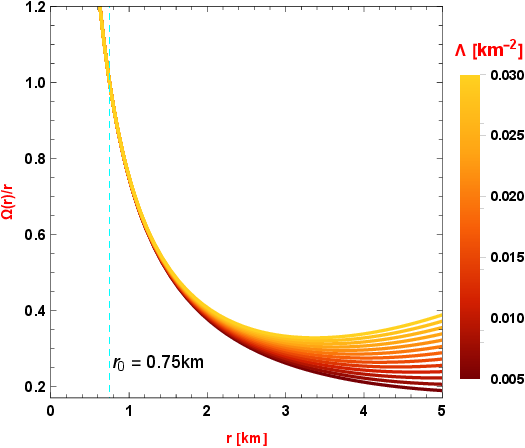}
			\includegraphics[width=5.6cm]{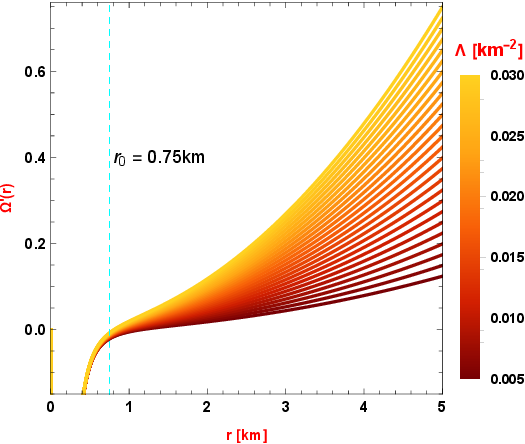}
			\\
		\end{tabular}
	\end{center}
	\caption{\label{fig3} Depiction of the shape function $\Omega(r)$ (Left), $\Omega(r)/r$ (Middle), $\Omega'(r)$ (Right) against the radial coordinate $r$ for the energy density model-II with parameters $r_0 = 0.75 km$, $\alpha = -0.8 km$, $S_0 = 0.01 km^{-1}$, $\lambda = 1.45km^{-1}$, $\mu = 1 km^{-1}$, and $\nu = 1$.}
\end{figure}

Also, the spin part of the conservation equation can be treated as being satisfied independently, which leads to
\begin{eqnarray}
    \Phi'(r)S^2+\frac{1}{2}(S^2)'=0.\label{sp}
\end{eqnarray}
Thus, we obtain
\begin{eqnarray}
    S^2=S_0^2 \exp(-2\Phi(r)).
\end{eqnarray}

Here, $S_0$ denotes an integration constant with dimension of $length^{-1}$. To construct explicit wormhole solutions, we choose the redshift function in the form $\Phi(r) = \frac{\alpha}{r}$, where $\alpha$ is a constant with dimension of $length$. This particular choice is motivated by its mathematical simplicity and physical viability. Importantly, the redshift function $\Phi(r) = \frac{\alpha}{r}$ remains finite for all $r> 0$, thereby avoiding divergences or singular behavior throughout the wormhole spacetime. Consequently, the absence of singularities ensures that no event horizon forms, which is a requirement for maintaining the traversability of the wormhole.

\section{Energy Conditions }\label{sec4}

In the study of wormholes, energy conditions serve as constraints on the stress–energy tensor $T_{\mu\nu}$, encapsulating fundamental expectations about the behavior of matter and energy. Within the framework of general relativity, sustaining a traversable wormhole inevitably entails the violation of the null energy condition (NEC) \cite{MT88, ms88a}. In contrast, the modified theories of gravity can respect or evade this requirement due to their effective field equations differ from those of general relativity. The literature suggests that the energy conditions originate naturally from the Raychaudhuri equations, which describe the temporal evolution of congruences of timelike vectors $u^\eta$ and the null geodesics $k_\eta$. These equations take the following form \cite{ar55}
\begin{eqnarray}
     &&\frac{d\Theta}{d\tau}-\omega_{\eta\xi}\omega^{\eta\xi}+\sigma_{\eta\xi}\sigma^{\eta\xi}+\frac{1}{3}\theta^2+R_{\eta\xi}u^\eta u_\xi = 0,
     \\
     &&\frac{d\Theta}{d\tau}-\omega_{\eta\xi}\omega^{\eta\xi}+\sigma_{\eta\xi}\sigma^{\eta\xi}+\frac{1}{3}\theta^2+R_{\eta\xi}k^\eta k_\xi = 0,
\end{eqnarray}
where $k^\eta$ stands for the vector files, $R_{\eta\xi}k^\eta k_\xi$ is the shear or spatial tensor with $\sigma ^2 = \sigma_{\eta\xi}\sigma^{\eta\xi} \geq 0$ and $\omega_{\eta\xi}\equiv 0$. For the case of attractive gravity ($\theta < 0$), the Raychaudhuri equations imply the following conditions
\begin{eqnarray}
   R_{\eta\xi}u^\eta u_\xi \geq 0,
   \\
   R_{\eta\xi}k^\eta k_\xi \geq 0.
\end{eqnarray}

\begin{figure}[h]
	\begin{center}
		\begin{tabular}{rl}
			\includegraphics[width=6cm]{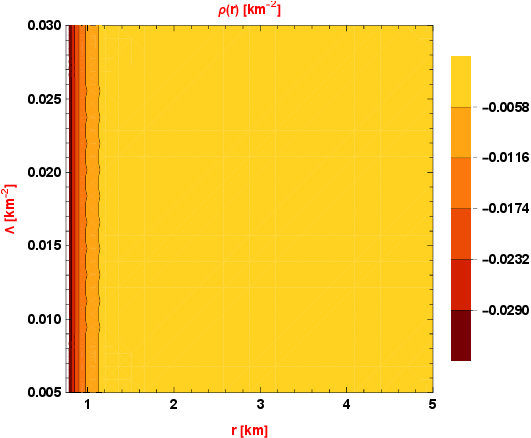}
			\includegraphics[width=5.7cm]{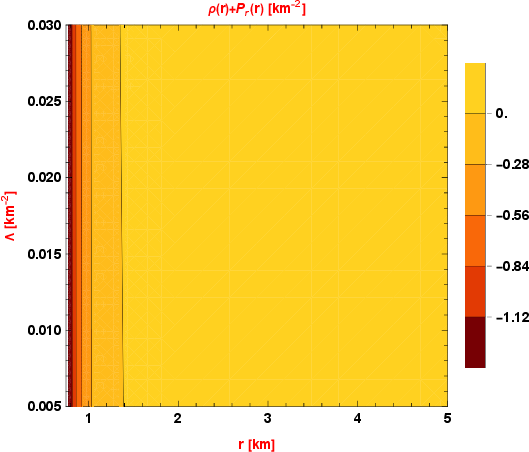}
			\includegraphics[width=5.6cm]{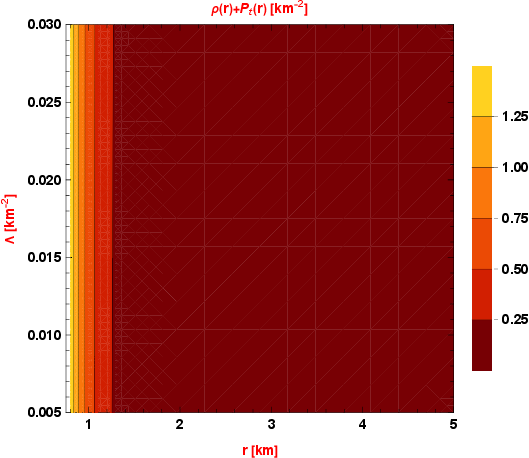}
			\\
		\end{tabular}
		\begin{tabular}{rl}
			\includegraphics[width=5.9cm]{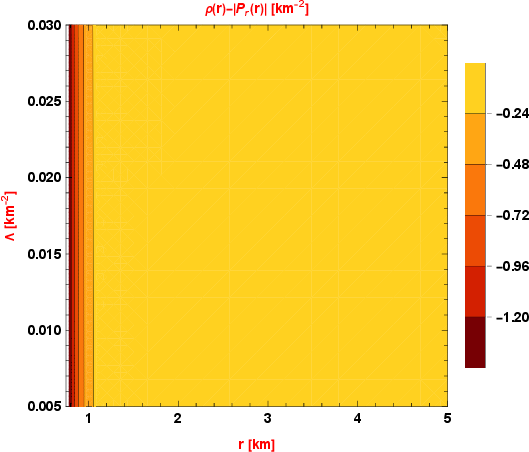}
			\includegraphics[width=5.8cm]{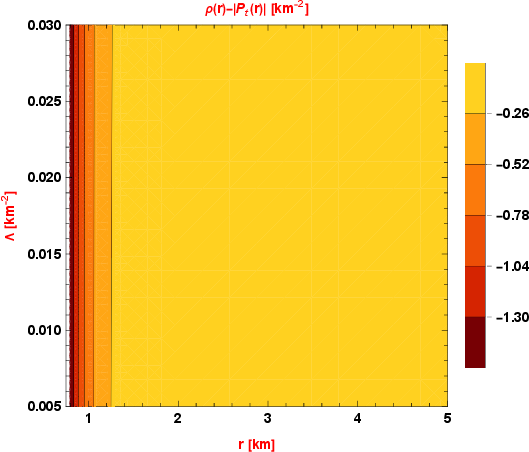}
			\includegraphics[width=5.6cm]{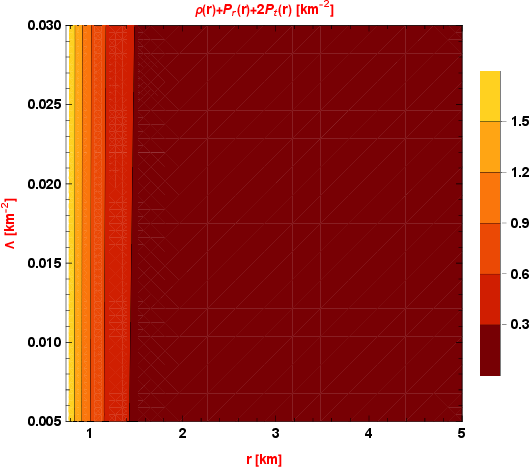}
			\\
		\end{tabular}
	\end{center}
	\caption{\label{fig4} Depiction of the energy density $\rho(r)$ (Left), $\rho(r)+P_r(r)$ (Middle), $\rho(r)+P_t(r)$ (Right) in the above panel, and  $\rho(r)-|P_r(r)|$ (Left), $\rho(r)-|P_t(r)|$ (Middle), $\rho(r)+P_r(r)+2P_t(r)$ (Right) in the below panel for the energy density model-II with parameters $r_0 = 0.75 km$, $\alpha = -0.8 km$, $S_0 = 0.01 km^{-1}$, $\lambda = 1.45km^{-1}$, $\mu = 1 km^{-1}$, and $\nu = 1$.}
\end{figure}

Thus, the energy conditions for an anisotropic matter configuration can be formulated as
\begin{itemize}
\item Null energy condition (NEC): $\rho(r) + P_r(r) \geq 0,~ \rho(r) + P_t(r) \geq 0$.
\item Weak energy condition (WEC): $\rho(r) \geq 0,~\rho(r) + P_r(r) \geq 0,~ \rho(r) + P_t(r) \geq 0$.
 \item Dominant energy condition (DEC): $\rho(r) \geq 0,~\rho(r) - |P_r(r)| \geq 0,~ \rho(r) - |P_t(r)| \geq 0$.
\item Strong energy condition (SEC): $\rho(r) + P_r(r) \geq 0, ~ \rho(r) + P_t(r) \geq 0, ~ \rho(r) + P_r(r)+2P_t(r) \geq 0$.
\end{itemize}

\section{Yukawa-Casimir Energy Density}\label{sec5}
In 1935, Yukawa \cite{hy35} introduced a seminal model to describe the strong nuclear interaction between nucleons within a nonrelativistic framework. He postulated that the nuclear force arises from the exchange of an intermediate massive field (later identified as the pion), which mediates the interaction in a manner analogous to how the photon mediates electromagnetic interactions. This theoretical insight not only provided a fundamental explanation for the short-range nature of the nuclear force but also laid the groundwork for modern quantum field theory approaches to particle interactions. The corresponding potential, now known as the Yukawa potential, takes the form
\begin{eqnarray}
     \mathcal{V}(r)=-\frac{\chi}{r}e^{-\lambda r},\label{vv}
\end{eqnarray}

where $\chi$ represents the strength of the nucleon interaction, while its range is characterized by $\frac{1}{\lambda}$. Several researchers have utilized this short-range interaction to explore possible deviations from the Newtonian potential. If such deviations exist, the Newtonian gravitational potential would acquire a Yukawa-type correction, which can be written in a form analogous to Eq. (\ref{vv}) as 
\begin{eqnarray}
     \mathcal{V}(r)=-\frac{\mathcal{G} m_1m_2}{r}\left(1+\chi e^{-\lambda r}\right),\label{vf}
 \end{eqnarray}
 
where $m_1$ and $m_2$ are two point masses separated by a distance $r$. It is noted that the potentials (\ref{vf}) have been widely investigated in astrophysical contexts, with particular attention to their implications for a finite graviton mass \cite{bd13, af16, af18}. Moreover, Yukawa-type forces are also anticipated within the framework of several modified gravity theories \cite{sc20, id18, md18}. In this context, the possibility of constructing traversable wormholes also remains viable in various theoretical frameworks \cite{jw15, aj22}.

In this study, we consider the Casimir energy density modified by the Yukawa-type term, referred to as the Yukawa-Casimir energy density. This modification gives rise to three distinct forms of Yukawa-Casimir energy density models, described as follows \cite{rg21}
\begin{eqnarray*}
    \text{Energy Density Model-1:} \quad\quad\quad \rho_y &=& r_0\rho_c \frac{e^{\lambda (r_0-r)}}{r},
    \\
   \text{Energy Density Model-2:} \quad\quad\quad \rho_y &=& \frac{\rho_c}{2r} \left(\mu r+\nu r_0e^{\lambda (r_0-r)}\right),
   \\
   \text{Energy Density Model-3:} \quad\quad\quad \rho_y &=& \frac{\rho_c r_0}{r} \left((1-\nu)e^{\lambda (r_0-r)}-\nu e^{\beta (r_0-r)}\right),
\end{eqnarray*}
 where $\rho_c$ is the original Casimir energy density, defined as \cite{mb15} 
\begin{eqnarray}
    \rho_c &=&-\frac{\pi ^2}{720 r^4}.
\end{eqnarray}

and $\lambda$, $\mu$, $\nu$, $\beta$ are model parameters that control the amplitude and decay of the energy density profiles. It is important to note that $\lambda,\mu,\beta$ have dimensions of $length^{-1}$, while $\nu$ is dimensionless.  Each model provides a phenomenological way to incorporate both quantum Casimir effects and short-range Yukawa-type interactions in the wormhole matter distribution. A brief summary of these models is provided in the following Table-\ref{YC}: 

\begin{table}[h!]
\centering
\caption{Summary of Yukawa-Casimir Energy-Density Models}\label{YC}
\begin{tabular}{|c|c|c|}
\hline
 \textbf{Energy Density Model} & \textbf{Physical Motivation / Interpretation} \\
\hline
 $\rho_y = r_0 \rho_c \dfrac{e^{\lambda (r_0 - r)}}{r}$ & Simple Yukawa-modified Casimir energy; negative energy near throat from   \\
& quantum vacuum fluctuations; short-range Yukawa interaction. \\
\hline
 $\rho_y = \dfrac{\rho_c}{2r} \Big(\mu r + \nu r_0 e^{\lambda (r_0 - r)} \Big)$ & Combines a slowly varying component with Yukawa-Casimir term; allows \\
& long-range background. \\
\hline
 $\rho_y = \dfrac{\rho_c r_0}{r} \Big( (1-\nu) e^{\lambda (r_0 - r)} - \nu e^{\beta (r_0 - r)} \Big)$ & Superposition of two Yukawa-like terms with different decay scales; models\\
&  more complex vacuum polarization effects and multiple interaction ranges.\\
\hline
\end{tabular}
\end{table}

Traversable wormholes generally require matter sources that effectively violate the NEC near the throat. In the Einstein–Cartan gravity framework, spacetime torsion provides additional degrees of freedom that naturally couple to the intrinsic spin of matter, which can reduce the amount of exotic matter needed to sustain a wormhole. In this context, Yukawa-Casimir energy-density profiles offer a physically motivated way to model the matter supporting the wormhole. The Casimir contribution arises from quantum vacuum fluctuations between boundaries or confining geometries, providing a local negative energy density that helps violate the NEC near the throat. The Yukawa term models short-range interactions mediated by massive fields, characterized by an interaction range determined by the inverse mass of the mediator, allowing the energy density to decay rapidly away from the throat. The combination of these two contributions produces a matter distribution that is negative near the throat, supports the traversable wormhole, while decaying sufficiently fast at large distances to remain compatible with the Einstein–Cartan field equations. By appropriately choosing the range parameters, the effective stress-energy tensor remains physically viable, providing a natural and theoretically motivated framework for exploring traversable wormholes supported by spin-torsion effects.

\begin{figure}[h]
\begin{center}
\begin{tabular}{rl}
\includegraphics[width=5.6cm]{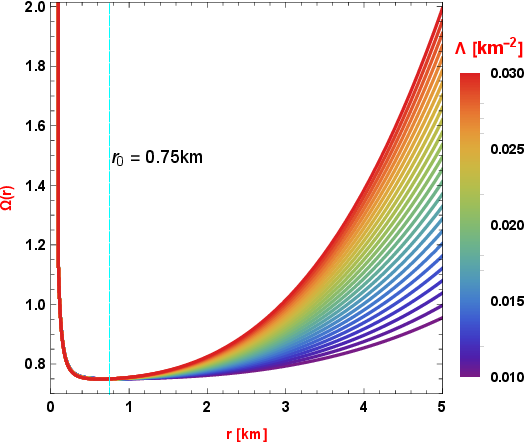}
\includegraphics[width=5.6cm]{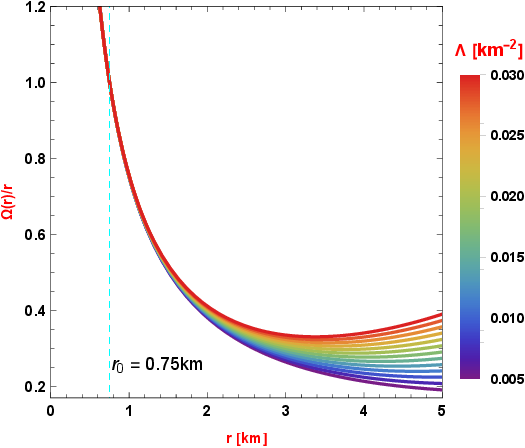}
\includegraphics[width=5.6cm]{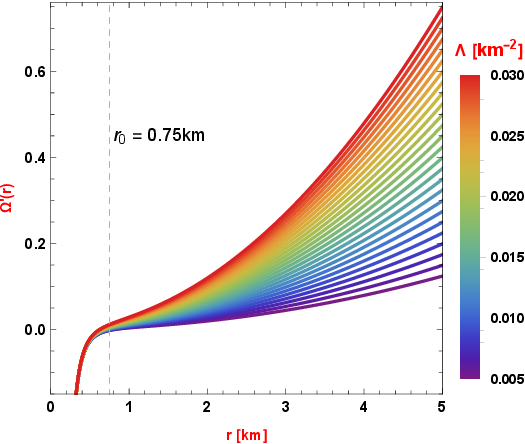}
\\
\end{tabular}
\end{center}
\caption{\label{fig5} Depiction of the shape function $\Omega(r)$ (Left), $\Omega(r)/r$ (Middle), $\Omega'(r)$ (Right) against the radial coordinate $r$ for the energy density model-III with parameters $r_0 = 0.75 km$, $\alpha = -0.8 km$, $S_0 = 0.01 km^{-1}$, $\lambda = 1.45km^{-1}$, $\nu$ =0.4, and $\beta = 1km^{-1}$.}
\end{figure}

\section{Yukawa-Casimir wormholes in Einstein-Cartan Gravity}\label{sec6}
In this section, we are willing to investigate the Yukawa-Casimir wormhole solutions supported by the aforementioned Yukawa-Casimir density profiles within the framework of EC gravity.

\subsection{Energy Density Model$-$1}\label{sec6a}

To investigate the wormhole solutions within the EC gravity, we first consider the following Yukawa-Casimir energy density \cite{rg21}:

\begin{eqnarray}
    \rho_y &=& r_0\rho_c \frac{e^{\lambda (r_0-r)}}{r},\label{rho1}
\end{eqnarray}

where $\lambda$ is the positive mass scale parameter. It should be noted that for $r=r_0$ the above equation reduces to the Casimir energy density $\rho_c$. Now, substituting the Yukawa$-$Casimir energy density (\ref{rho1}) into the field Eq. (\ref{den}), we obtain 
\begin{eqnarray}
    r_0\rho_c \frac{e^{\lambda (r_0-r)}}{r} = \frac{1}{4r^2}\left[4\Omega^\prime(r)+r^2 S^2(r)-4\Lambda r^2\right].
\end{eqnarray}
After solving the above differential equation, the shape function $\Omega(r)$ is obtained as
\begin{eqnarray}
    \Omega(r) = \frac{1}{720} \left[\pi^2 \lambda r_0 \mathcal{H}(r) e^{\lambda r_0}+\frac{\pi^2 r_0}{r} e^{\lambda(r_0-r)}-240 \alpha ^3 S_0^2 \mathcal{K}(r)-60 S_0^2 r e^{-\frac{2 \alpha }{r}} \left(r^2-2 \alpha  r+8 \alpha ^2\right)+240 \left(3 \mathcal{C}_1+\Lambda r^3\right)\right],
\end{eqnarray}
where $\mathcal{H}(r) = \text{ExpIntegralEi}[-\lambda r]$, $\mathcal{K}(r) = \text{ExpIntegralEi}[-2\alpha/ r]$, and $\mathcal{C}_1$ is an integration constant. Here, we impose the throat condition $\Omega(r_0) = r_0$ to determine the constant $\mathcal{C}_1$, this yields the following values for $\mathcal{C}_1$ 
\begin{eqnarray}
    \mathcal{C}_1 = \frac{1}{720} \left[60 r_0 \left\{S_0^2 e^{-\frac{2 \alpha }{r_0}} \left(8 \alpha ^2+r_0^2-2 \alpha  r_0\right)-4\Lambda r_0^2+12\right\}-\pi^2 \lambda r_0 \mathcal{H}(r_0)  e^{\lambda r_0}+240\alpha^3 S_0^2 \mathcal{K}(r_0)-\pi^2\right].
\end{eqnarray}

Accordingly, the resulting expression for the shape function is obtained as 
\begin{eqnarray}
\Omega(r)&=&  \frac{1}{720} \Big[\pi^2 \lambda r_0 e^{\lambda r_0} (\mathcal{H}(r) -\mathcal{H}(r_0))+\frac{\pi^2 r_0}{r} e^{\lambda(r_0-r)}-240\alpha ^3 S_0^2 (\mathcal{K}(r)-\mathcal{K}(r_0))-60 S_0^2 r e^{-\frac{2\alpha }{r}} \left(r^2-2 \alpha  r+8 \alpha ^2\right)\nonumber
\\
&&+240\Lambda (r^3-r_0^3)+60 r_0S_0^2 e^{-\frac{2 \alpha }{r_0}} \left(r_0^2-2 \alpha  r_0+8\alpha^2\right)+720 r_0-\pi^2\Big].\label{B1}
\end{eqnarray}

In addition, the wormhole solutions must satisfy the flare–out condition at the wormhole throat, which is characterized by the relation 
\begin{eqnarray}
\Omega'(r_0)&=& \Lambda r_0^2-\frac{r_0^2}{4} S_0^2 e^{-\frac{2\alpha }{r_0}}-\frac{\pi ^2}{720 r_0}  < 1. \label{b1}
\end{eqnarray}

To gain deeper insight into the properties of the derived shape function (\ref{B1}), we investigate its behavior using graphical illustrations. In this regards, we present the plots of  $\Omega(r)$, $\Omega(r)/r$ and $\Omega'(r)$ in Fig. \ref{fig1} for representative parameter choices $r_0 = 0.75km$, $\alpha = -0.8km$, $S_0 = 0.01km^{-1}$, $\lambda = 1.45km^{-1}$, and $\Lambda \in [0.005km^{-2}, 0.03km^{-2}]$. From Fig. \ref{fig1}, it is observed that the shape function $\Omega(r)$ increases monotonically with respect to both the radial coordinate $r$ and the parameter $\Lambda$, while consistently satisfying the condition $\Omega(r)/r \leq 1$ for $r \geq r_{0}$. In addition, the shape function is found to comply with the flare-out condition under this configuration. These findings allow us to conclude that the newly derived shape function supported by the Yukawa-Casimir energy density (\ref{rho1}) successfully encapsulates all the essential features required for traversable wormhole geometries.  It is important to emphasize that, in the present case, the shape function fails to exhibit the desired asymptotic behavior. Consequently, the proposed wormhole geometry must be truncated at a finite radius $r = \mathcal{R}_1$ and smoothly matched to an exterior Schwarzschild solution.

The expressions for the radial and transverse pressures can now be derived from the field equations (\ref{pr}) and (\ref{pt}), yielding the following forms: 
\begin{eqnarray}
    P_r(r) &=& \frac{1}{720 r^5}\left[r\xi_1-\pi^2 r_0(r-2\alpha) e^{\lambda(r_0-r)}+60 S_0^2 r e^{-2 \alpha  (r_0+r)/r_0 r}\xi_2+r (r-2\alpha) \xi_3\right],
    \\
    P_t(r) &=& \frac{e^{-2 \alpha(r+r_0)/r_0r}}{1440 r^6}\Big[60r_0^3 re^{\frac{2\alpha }{r}} \left(r^2-3 \alpha r-2\alpha^2\right) \left(S_0^2-4\Lambda e^{\frac{2\alpha}{r_0}}\right)-60\alpha r_0^2 S_0^2re^{\frac{2\alpha}{r}}\left(r^2-3\alpha r-2\alpha^2\right)\nonumber
    \\
    &&- r \left(-2 \alpha ^2+r^2-3 \alpha  r\right) e^{\alpha(r+r_0)/r_0 r}\xi_3+2r_0 e^{\frac{2\alpha }{r}}\xi_4+re^{\frac{2\alpha }{r_0}}\xi_5\Big],
\end{eqnarray}
where
\begin{eqnarray}
    \xi_1 &=& 240\Lambda r_0^3 (r-2\alpha)-720 r_0(r-2\alpha )+480 \left(\Lambda r^4+\Lambda \alpha r^3-3\alpha r\right)+\pi^2(r-2\alpha ),\nonumber
    \\
    \xi_2 &=& r e^{\frac{2 \alpha }{r_0}} \left(4r^3-3\alpha r^2+4\alpha^2r-4\alpha^3\right)-r_0\left(2\alpha^2+r_0^2-\alpha r_0\right) e^{\frac{2\alpha }{r}} (r-2\alpha ),\nonumber
    \\
    \xi_3 &=& \pi^2\lambda r_0e^{\lambda r_0} (\mathcal{H}(r_0)-\mathcal{H}(r))+240 \alpha ^3 k^2 (\mathcal{K}(r)-\mathcal{K}(r_0)),
    \\
    \xi_4 &=& 60r\left(r^2-3 \alpha r-2 \alpha^2\right) \left(6 e^{\frac{2\alpha }{r_0}}+\alpha^2 S_0^2\right)+\pi^2\left(r^2-2 \alpha r-\alpha^2\right) e^{\frac{2\alpha}{r_0}+\lambda r_0 -\lambda r},
    \\
    \xi_5 &=& e^{\frac{2\alpha}{r}} \left[\pi^2 \left(2\alpha^2-r^2+3\alpha r\right)+480r \left(2\Lambda r^4+\alpha^2 \left(3-\Lambda r^2 \right)+3\alpha r\right)\right]+60S_0^2 r \big[8 r^4+\alpha  r^3-3 \alpha ^2 r^2\nonumber
    \\
    &&+4 \alpha ^3 r+4 \alpha ^4\big].
\end{eqnarray}

Furthermore, the explicit expression for the NEC evaluated at the wormhole throat can be written as 
\begin{equation}
\left[\rho(r)+P_{r}(r)\right]_{r=r_{0}}=\Lambda+\frac{S_{0}^{2}}{4}e^{-\frac{2\alpha}{r_0}}-\frac{720r_{0}+\pi^{2}}{720r_{0}^{3}}.    
\end{equation}

To assess the physical nature of the matter threading the wormhole structures, we provide detailed graphical illustrations of all relevant energy conditions, as displayed in Fig. \ref{fig2} corresponding to $r_0 = 0.75km$, $\alpha = -0.8km$, $S_0 = 0.01km^{-1}$, $\lambda = 1.45km^{-1}$, and $\Lambda \in [0.005km^{-2}, 0.03km^{-2}]$. It is observed that the Yukawa-Casimir energy density consistently preserves its negative character across the full spacetime domain. Moreover, the NEC, DEC, and SEC are found to be violated near the wormhole throat, indicating the presence of exotic matter within the wormhole structures, which plays a crucial role in sustaining the constructed wormhole geometries in the framework of EC gravity.

\begin{figure}[h]
	\begin{center}
		\begin{tabular}{rl}
			\includegraphics[width=6cm]{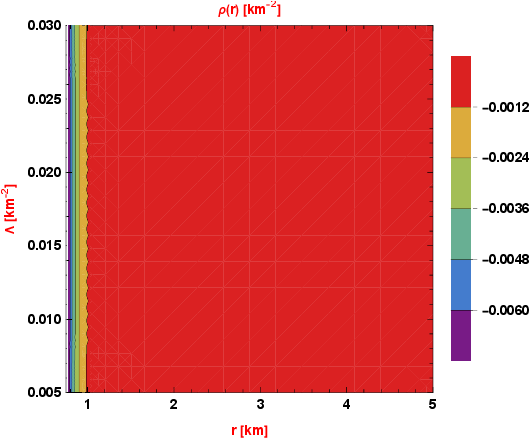}
			\includegraphics[width=5.8cm]{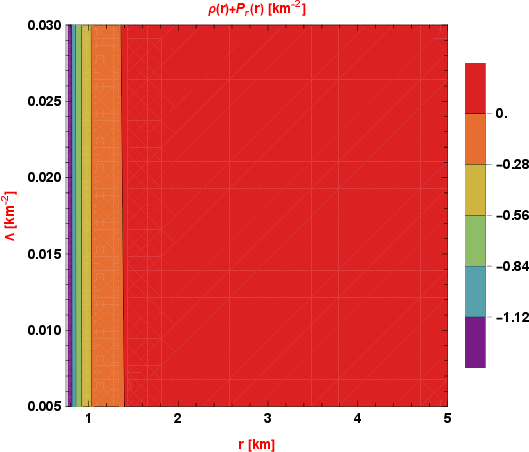}
			\includegraphics[width=5.6cm]{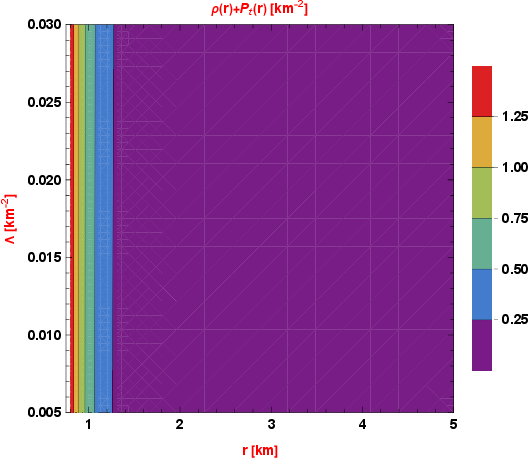}
			\\
		\end{tabular}
		\begin{tabular}{rl}
			\includegraphics[width=5.9cm]{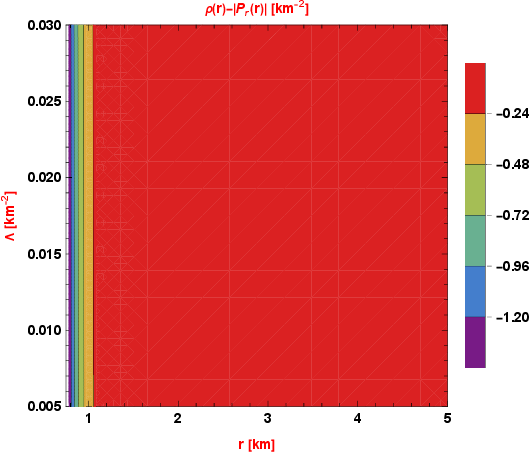}
			\includegraphics[width=5.8cm]{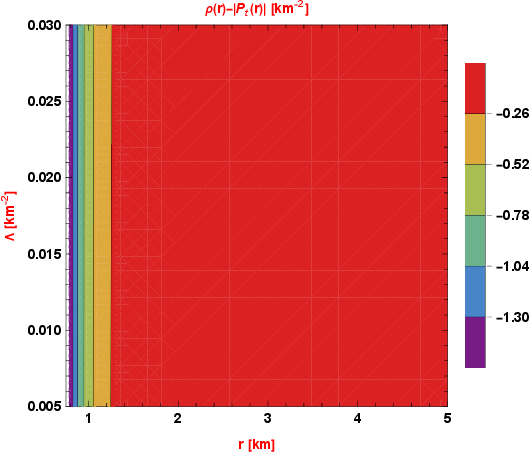}
			\includegraphics[width=5.6cm]{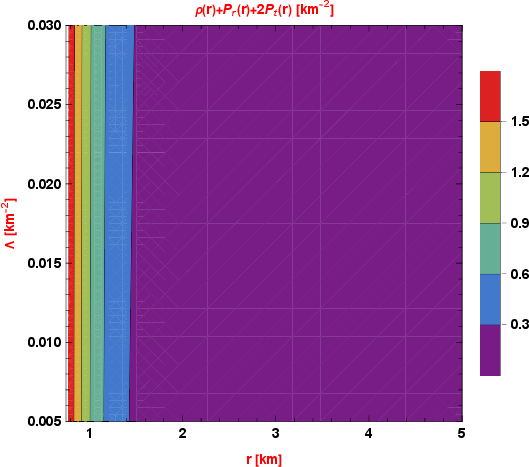}
			\\
		\end{tabular}
	\end{center}
	\caption{\label{fig6} Depiction of the energy density $\rho(r)$ (Left), $\rho(r)+P_r(r)$ (Middle), $\rho(r)+P_t(r)$ (Right) in the above panel, and  $\rho(r)-|P_r(r)|$ (Left), $\rho(r)-|P_t(r)|$ (Middle), $\rho(r)+P_r(r)+2P_t(r)$ (Right) in the below panel for the energy density model-III with parameters $r_0 = 0.75 km$, $\alpha = -0.8 km$, $S_0 = 0.01 km^{-1}$, $\lambda = 1.45km^{-1}$, $\nu$ =0.4, and $\beta = 1km^{-1}$.}
\end{figure}

\subsection{Energy Density Model$-$2}\label{sec6b}
In this section, we are going to consider the second Yukawa-Casimir energy density to investigate the wormhole solutions within EC gravity, expressed as
\begin{eqnarray}
\rho_y &=& \frac{\rho_c}{2r} \left(\mu r+\nu r_0e^{\lambda (r_0-r)}\right),\label{rho2}
\end{eqnarray}
where $\mu,~\nu~\in~ \mathbb{R}$. Indeed, the above density model is a linear combination of the original Casimir profile and the Yukawa-Casimir profile (\ref{rho1}). Notably, for $\mu = \nu = 1$ and $r=r_0$, the above density profile (\ref{rho2}) reduces to the original Casimir energy density $\rho_c$. 

The shape function in this case is derived by using the Yukawa energy density (\ref{rho2}) in the field Eq. (\ref{den}), yielding 
\begin{eqnarray}
    \Omega(r)&=&\frac{1}{2880r^2}\Big[\pi^2\nu r_0e^{\lambda(r_0-r)} \left(1-\lambda r-\mathcal{H}(r) \lambda^2r^2e^{\lambda r}\right)-960\alpha^3 S_0^2 r^2\mathcal{K}(r)-240S_0^2 r^3 e^{-\frac{2 \alpha }{r}} \left(r^2-\alpha r+2\alpha^2\right)\nonumber
    \\
    &&+960\Lambda r^5+2\pi^2\mu r\Big]+\mathcal{C}_2,
\end{eqnarray}
where $\mathcal{C}_2$ is an integration constant, which can be fixed using the throat condition $\Omega(r_0) = r_0$, leading to the following result 
\begin{eqnarray}
    \mathcal{C}_2 = \frac{r_0^2 \left[240S_0^2 e^{-\frac{2\alpha}{r_0}} \left(r_0^2-\alpha  r_0+2\alpha^2\right)-960\Lambda r_0^2+2880\right]+\pi^2\nu\left(\lambda^2 r_0^2 \mathcal{H}(r_0) e^{\lambda r_0}+\lambda r_0-1\right)+960\alpha^3 r_0 S_0^2\mathcal{K}(r_0)-2\pi^2\mu}{2880 r_0}.
\end{eqnarray}

Thus, the shape function takes the final form 
\begin{eqnarray}
\Omega(r)&=&\frac{\pi^2\nu r_0^2(1-\lambda r) e^{\lambda(r_0-r)}-240r_0S_0^2 r^3 e^{-\frac{2 \alpha }{r}} \left(r^2-\alpha r+2\alpha^2\right)+960\Lambda r_0 r^5+r^2\left(\xi_6+\xi_7\right)+2\pi^2\mu r_0r}{2880r_0r^2},\label{B2}
\end{eqnarray}
where
\begin{eqnarray}
    \xi_6 &=& r_0 \left[240 r_0 \left(S_0^2 e^{-\frac{2 \alpha }{r_0}} \left(r_0^2-\alpha  r_0+2\alpha^2\right)-4\Lambda r_0^2+12\right)+\pi^2\lambda\nu\right]-\pi^2(2\mu +\nu), \nonumber
    \\
    \xi_7 &=& \pi^2\lambda^2\nu r_0^2e^{\lambda r_0} (\mathcal{H}(r_0)-\mathcal{H}(r))+960 \alpha^3S_0^2r_0(\mathcal{K}(r_0)-\mathcal{K}(r)).
\end{eqnarray}

In this case, the flare–out condition at the wormhole throat can be expressed by the following relation 
\begin{eqnarray}
\Omega'(r_0)&=& \Lambda r_0^2-\frac{r_0^2}{4} S_0^2 e^{-\frac{2 \alpha }{r_0}}-\frac{\pi ^2 (\mu +\nu )}{1440 r_0^2}  < 1. \label{b1}
\end{eqnarray}

The derived shape function (\ref{B2}) is analyzed  graphically in Fig.~\ref{fig3} with the parameters $r_0 = 0.75km$, $\alpha = -0.8km$, $S_0 = 0.01km^{-1}$, $\lambda = 1.45km^{-1}$, $\mu = 1km^{-1}$, $\nu = 1$, and $\Lambda \in [0.005km^{-2}, 0.03km^{-2}]$. Fig.~\ref{fig3}  shows that the shape function increases monotonically with both $r$ and $\Lambda$, satisfies $\Omega(r)/r \leq 1$ for $r \geq r_0$, and meets the flare-out condition. These results indicate that the shape function supported by the Yukawa-Casimir energy density (\ref{rho2})  captures the essential features of traversable wormhole geometries.  However, since it does not exhibit the correct asymptotic behavior, the wormhole  must be truncated at a finite radius $r = \mathcal{R}_1$ and matched to an 
exterior Schwarzschild solution.

The radial and transverse pressures for this case are obtained in the following forms 
\begin{eqnarray}
    P_r(r) &=& \frac{1}{2880r_0 r^6}\Big[r \big\{960 r_0 r \left(\Lambda r_0^3(r-2\alpha)-3r_0(r-2 \alpha )+2\Lambda r^3(\alpha +r)-6\alpha r\right)+240S_0^2 r_0r\xi_8e^{-2\alpha(r_0+r)/r_0 r}\nonumber
    \\
    &&-\pi^2\nu r (\lambda r_0 -1) (r-2\alpha )-2\pi^2\mu(r_0-r) (r-2\alpha)\big\}+\pi^2r_0^2\nu(r-2 \alpha) (\lambda r-1) e^{\lambda(r_0-r)}-r_0(r-2 \alpha ) \xi_9\Big].
    \\
    P_t(r) &=& \frac{1}{5760r_0 r^7}\Big[r_0\xi_9 \left(r^2-3\alpha r-2\alpha ^2\right)+r \Big\{240S_0^2 r_0^2r\left(r_0^2-\alpha r_0+2\alpha^2\right) e^{\frac{2\alpha}{r}} \left(r^2-3\alpha r-2\alpha^2\right)+e^{\frac{2\alpha}{r_0}} \Big(240S_0^2\times\nonumber
    \\
    &&r_0r^2(8r^4+\alpha r^3-3\alpha^2r^2+4\alpha^3r+4\alpha^4)+\xi_{10} e^{\frac{2\alpha}{r}}\Big)\Big\}e^{-2\alpha(r_0+r)/r_0r}+r_0^2\xi_{11}\Big],
\end{eqnarray}
where
\begin{eqnarray}
    \xi_8 &=& r e^{\frac{2\alpha }{r_0}} \left(4r^3-3\alpha r^2+4\alpha^2 r-4\alpha^3\right)-r_0(r-2 \alpha)\left(r_0^2-\alpha r_0+2\alpha^2\right) e^{\frac{2\alpha}{r}},\nonumber
    \\
    \xi_9 &=& r^2 \left[\pi^2\lambda^2\nu r_0 e^{\lambda r_0} (\mathcal{H}(r_0)-\mathcal{H}(r))+960\alpha^3 S_0^2 (\mathcal{K}(r_0)-\mathcal{K}(r))\right],\nonumber
    \\
    \xi_{10} &=&960r_0 r\left[2\alpha^2\left(\Lambda (r_0^3-r^3)-3(r_0-r)\right)-r^2\left(\Lambda r_0^3-3r_0-4\Lambda r^3\right)+3\alpha r\left(\Lambda r_0^3-3r_0+2r\right)\right]+\pi^2\nu r(\lambda r_0 -1)\big(r^2\nonumber
    \\
    &&-3 \alpha  r-2 \alpha ^2\big)+2\pi^2\mu \left[2r_0 \left(r^2-2\alpha r-\alpha^2\right)+r \left(2\alpha^2-r^2+3 \alpha  r\right)\right],\nonumber
    \\
   \xi_{11} &=& \pi ^2  \nu  \left[3(\alpha\lambda +1)r^2-2\alpha^2-\lambda r^3+\alpha(2\alpha\lambda-5)r\right] e^{\lambda (r_0-r)}.
\end{eqnarray}
In addition, upon evaluation at the throat $r = r_0$, the NEC is expressed as
\begin{eqnarray}
    \left[\rho(r)+P_r(r)\right]_{r=r_0} &=&\Lambda+\frac{S_0^2}{4}e^{-\frac{2\alpha}{r_0}}-\frac{1440 r_0^2+\pi^2 (\mu +\nu )}{1440r_0^4}.
\end{eqnarray}

The relevant energy conditions are depicted in Figure~\ref{fig4}, which shows that the Yukawa-Casimir energy density (\ref{rho2}) is negative throughout the spacetime. The NEC, DEC, and SEC are violated near the throat, indicating the presence of exotic matter that helps to sustain the wormhole within EC gravity.

\subsection{Energy Density Model$-$3}\label{sec6c}
Here, we consider another form of the Yukawa$-$Casimir energy density to generate the wormhole solutions in EC gravity, expressed as
\begin{eqnarray}
   \rho_y &=&\frac{\rho_c r_0}{r} \left((1-\nu)e^{\lambda (r_0-r)}-\nu e^{\beta (r_0-r)}\right),\label{rho3}
\end{eqnarray}
where $\lambda > 0$, $\beta > 0$. It is important to note that the above density profile (\ref{rho3}) reproduces both Yukawa behaviour and a generalized absurdly benign traversable wormhole (GABTW) \cite{rg20}. When $\nu = 0$, the density profile (\ref{rho3}) reduces to the density profile  (\ref{rho1}), while for $\nu = 1$ we get its replusive version. Therefore, $\forall~\nu \neq 0, 1$, we have a linear superposition of the Yukawa-Casimir profile. Moreover, for $\nu = 0$ and $r=r_0$, the density profile (\ref{rho3}) reduces to the original Casimir energy density $\rho_c$.

\begin{figure}[h]
	\begin{center}
		\begin{tabular}{rl}
			\includegraphics[width=5.9cm]{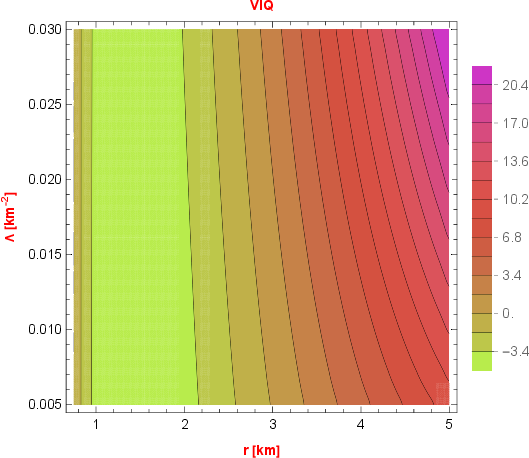}
			\includegraphics[width=5.9cm]{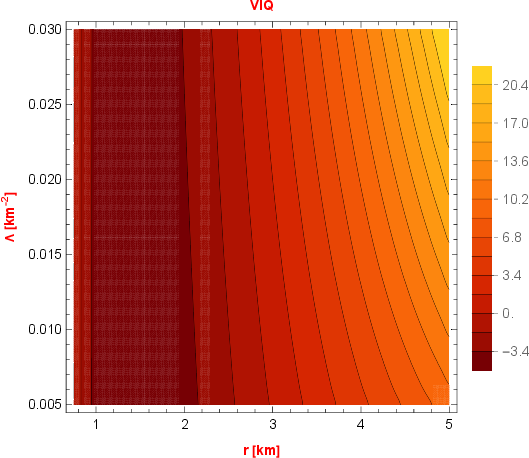}
			\includegraphics[width=5.8cm]{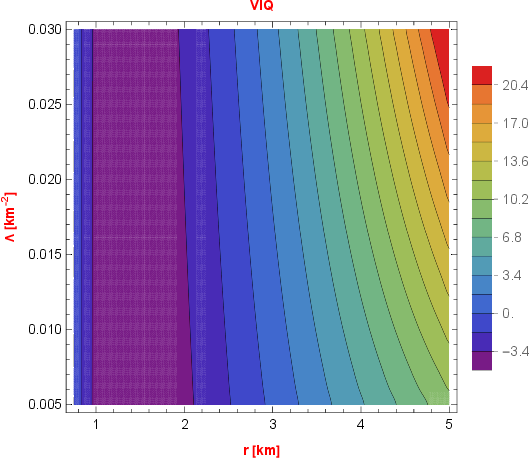}
			\\
		\end{tabular}
	\end{center}
	\caption{\label{fig7} Depiction of the $\mathcal{VIQ}$ for the energy density model-1 with $r_0 = 0.75 km$, $\alpha = -0.8 km$, $S_0 = 0.01 km^{-1}$, and $\lambda = 1.45km^{-1}$ (Left), for the energy density model-2 with $r_0 = 0.75 km$, $\alpha = -0.8 km$, $S_0 = 0.01 km^{-1}$, $\lambda = 1.45km^{-1}$, $\mu = 1 km^{-1}$, and $\nu = 1$ (Middle), and for the energy density model-3 with $r_0 = 0.75 km$, $\alpha = -0.8 km$, $S_0 = 0.01 km^{-1}$, $\lambda = 1.45km^{-1}$, $\nu$ =0.4, and $\beta = 1km^{-1}$ (Right).}
\end{figure}

For the Yukawa-Casimir energy density (\ref{rho3}), we obtain the following shape function 
\begin{eqnarray}
    \Omega(r)&=& \frac{1}{1440r^2}\Big[1440 \mathcal{C}_3 r^2+\pi^2r_0 r^2 \left\{\lambda^2(\nu -1) e^{\lambda r_0}\mathcal{H}(r)+\beta^2\nu e^{\beta r_0}\mathcal{J}(r)\right\}-480\alpha^3 S_0^2r^2\mathcal{K}(r)+480\Lambda r^5\nonumber
    \\
    &&+\pi^2r_0 \left\{\nu  (\beta  r-1) e^{\beta  (d-r)}+(\nu -1) (\lambda r-1) e^{\lambda(r_0-r)}\right\}-120S_0^2r^3\left(r^2-\alpha r+2\alpha^2\right)e^{-\frac{2 \alpha}{r}}\Big],
\end{eqnarray}
where $\mathcal{C}_3$ is an integration constant and $\mathcal{J}(r) =  \text{ExpIntegralEi}[-\beta r]$. Similarly, the throat condition $\Omega(r_0) = r_0$ yields the result for $\mathcal{C}_3$ in the following form 
\begin{eqnarray}
    \mathcal{C}_3 &=& \frac{1}{1440r_0}\Big[120S_0^2r_0^2\left(r_0^2-\alpha  r_0+2\alpha^2\right)e^{-\frac{2\alpha }{r_0}}-480\Lambda r_0^4-\pi^2r_0^2 \left(\lambda^2(\nu -1) e^{\lambda r_0}\mathcal{H}(r_0)+\beta^2\nu e^{\beta r_0}\mathcal{J}(r_0)\right)+1440 r_0^2\nonumber
    \\
    &&-\pi^2 (\beta\nu r_0+\lambda(\nu -1)r_0-2 \nu +1)+480\alpha^3S_0^2r_0\mathcal{K}(r_0)\Big].
\end{eqnarray}

Finally, the shape function takes the following form 
\begin{eqnarray}
\Omega(r)&=&\frac{1}{1440r_0r^2}\Big[\pi^2 r_0^2 \xi_{12}+480\alpha^3S_0^2r_0r^2(\mathcal{K}(r_0)-\mathcal{K}(r))-120S_0^2r_0 r^3e^{-\frac{2\alpha }{r}}\left(2\alpha^2+r^2-\alpha  r\right)+480\Lambda r_0r^5 +r^2\xi_{13} \nonumber
\\
&&-\pi^2\lambda^2(\nu -1)r_0^2r^2e^{\lambda r_0} (\mathcal{H}(r_0)-\mathcal{H}(r))-\pi^2\beta^2\nu r_0^2r^2 e^{\beta r_0} (\mathcal{J}(r_0)-\mathcal{J}(r))\Big],\label{B3}
\end{eqnarray}
where
\begin{eqnarray}
    \xi_{12} &=&\nu(\beta r-1) e^{\beta(r_0-r)}+(\nu -1)(\lambda r-1) e^{\lambda(r_0-r)},\nonumber
    \\
    \xi_{13} &=& 120S_0^2r_0^2\left(r_0^2-\alpha r_0+2\alpha^2\right)e^{-\frac{2\alpha}{r_0}}-480\Lambda r_0^4+1440r_0^2-\pi^2\left[\nu(\beta +\lambda )r_0-\lambda r_0-2\nu +1\right].
\end{eqnarray}

Also, the flaring-out condition at the wormhole throat can be expressed as 
\begin{eqnarray}
\Omega'(r_0)&=&  \Lambda r_0^2-\frac{r_0^2}{4} S_0^2 e^{-\frac{2\alpha }{r_0}}+\frac{\pi ^2 (2 \nu -1)}{720 r_0^2}  < 1. \label{b1}
\end{eqnarray}

The shape function (\ref{B3}) is illustrated in Fig.~\ref{fig5} for  $r_0 = 0.75km$, $\alpha = -0.8km$, $S_0 = 0.01km^{-1}$, $\lambda = 1.45km^{-1}$, $\beta = 1km^{-1}$, $\nu = 0.4$, and $\Lambda \in [0.005km^{-2}, 0.03km^{-2}]$.  It is observed that the obtained shape function increases monotonically with both $r$ and $\Lambda$. Moreover, it satisfies the condition $\Omega(r)/r \leq 1$ for $r \geq r_0$, and fulfils the flare-out condition nicely for $r \geq r_0$. Therefore, this shape function also nicely satisfies the essential features of traversable wormhole geometries. Notably, this shape function also does not exhibit the asymptotic nature, therefore, the wormholes must be matched to an exterior Schwarzschild solution at a finite radius $r = \mathcal{R}_1$.

In this scenario, the radial pressure and transverse pressure are obtained as 
\begin{eqnarray}
    P_r(r) &=& \frac{1}{1440r_0 r^6}\Big[\pi^2r_0^2(2 \alpha-r) \left\{\nu(\beta r-1)e^{\beta  (r_0-r)}+(\nu -1)(\lambda r-1) e^{\lambda(r_0-r)}\right\}+r^2 \Big\{r_0 \big(2\pi^2\alpha\lambda +\pi^2\nu(\beta +\lambda)(r-2\alpha )\nonumber
    \\
    &&-\pi^2\lambda r\big)+\xi_{14}-\pi^2(2\nu -1)(r-2\alpha)\Big\}+r_0r^2(r-2\alpha)\xi_{15}\Big],
    \\
    P_t(r) &=& \frac{1}{2880r_0 r^7}\Big[\pi^2(2\nu -1)r^2\left(r^2-3\alpha r-2\alpha ^2\right)-r_0r^2 \left(r^2-3\alpha r-2\alpha ^2\right)\xi_{15}+\Big\{120r_0^4r^2e^{\frac{2\alpha}{r}} \left(r^2-3 \alpha r-2\alpha^2\right) \big(S_0^2\nonumber
    \\
    &&-4\Lambda e^{\frac{2\alpha}{r_0}}\big)-120 \alpha S_0^2r_0^3r^2\left(r^2-3\alpha r-2\alpha^2\right)e^{\frac{2\alpha }{r}}+r_0r^2 e^{\frac{2\alpha }{r_0}}\xi_{16}+r_0^2\xi_{17} e^{\frac{2\alpha }{r}-r(\beta+\lambda)}\Big\}e^{-2 \alpha(r_0+r)/r_0 r}\Big],
\end{eqnarray}
where
\begin{eqnarray}
    \xi_{14} &=& r_0\left[480\Lambda r_0^3(r-2 \alpha )-1440r_0(r-2\alpha)+960\Lambda r^4+960\Lambda\alpha r^3-2880\alpha r\right]+120S_0^2 r_0 e^{-2\alpha(r_0+r)/r_0 r}\Big[r e^{\frac{2\alpha}{r_0}}\big\{4r^3\nonumber
    \\
    &&-3\alpha r^2+4\alpha^2r-4\alpha^3\big\}-r_0\left(2\alpha^2+r_0^2-\alpha r_0\right)(r-2\alpha)e^{\frac{2\alpha }{r}}\Big],\nonumber
    \\
    \xi_{15} &=& \pi^2 r_0 \left[\lambda^2(\nu -1) e^{\lambda r_0} (\mathcal{H}(r_0)-\mathcal{H}(r))+\beta^2\nu e^{\beta r_0} (\mathcal{J}(r_0)-\mathcal{J}(r))\right]+480\alpha^3 S_0^2 (\mathcal{K}(r)-\mathcal{K}(r_0)),\nonumber
    \\
   \xi_{16} &=& 120 S_0^2 r \left(8r^4+\alpha r^3-3\alpha^2r^2+4\alpha^3r+4\alpha^4\right)+e^{\frac{2\alpha}{r}} \Big[2\pi^2\alpha^2(\nu  (\beta +\lambda)-\lambda)+1920\Lambda r^5-960\Lambda \alpha^2 r^3+r^2 \big\{2880 \alpha\nonumber
   \\
   && +\pi^2(\lambda-\nu(\beta +\lambda ))\big\}+3\alpha r\left\{960\alpha +\pi^2(\nu(\beta +\lambda)-\lambda)\right\}\Big],\nonumber
    \\
   \xi_{17} &=& 1440r^2\left(r^2-3\alpha r-2\alpha^2\right) e^{\frac{2\alpha }{r_0}+r(\beta +\lambda)}+\pi^2(\nu -1)\left(\lambda r^3-3r^2(\alpha \lambda +1)-\alpha r(2\alpha\lambda -5)+2\alpha^2\right) e^{\frac{2\alpha }{r_0}+\lambda r_0 +\beta r}\nonumber
   \\
   &&+\pi^2\nu\left(\beta r^3-3r^2(\alpha\beta +1)-\alpha r(2\alpha \beta -5)+2\alpha^2\right) e^{\frac{2\alpha }{r_0}+\beta r_0+\lambda r}+240\alpha^2 S_0^2 r^2 \left(r^2-3\alpha r-2 \alpha ^2\right) e^{r (\beta +\lambda )}.
\end{eqnarray}

In this case, evaluating the NEC at the throat $r = r_0$ yields 
\begin{eqnarray}
    \left[\rho(r)+P_r(r)\right]_{r=r_0} &=&\Lambda+\frac{S_0^2}{4}e^{-\frac{2\alpha}{r_0}}-\frac{720 r_0^2+\pi^2 (1-2\nu )}{720r_0^4}.
\end{eqnarray}

\begin{figure}[h]
	\begin{center}
		\begin{tabular}{rl}
			\includegraphics[width=5.7cm]{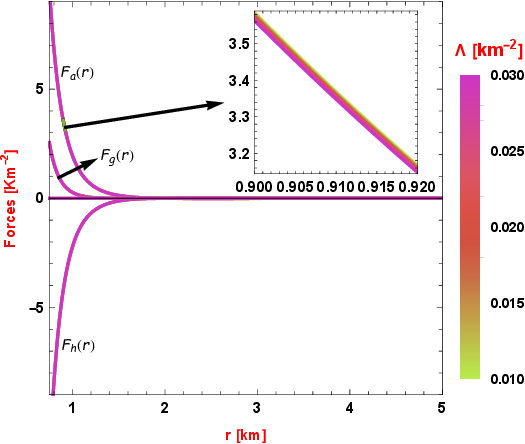}
			\includegraphics[width=5.7cm]{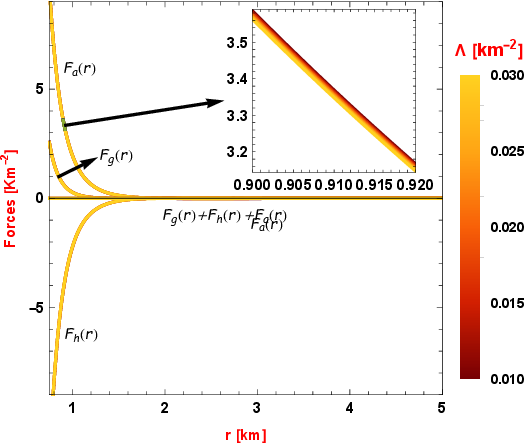}
			\includegraphics[width=5.7cm]{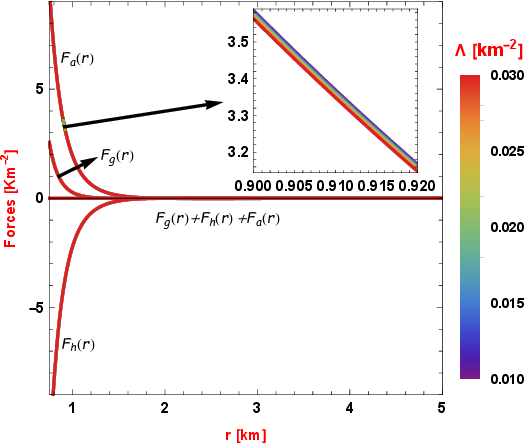}
			\\
		\end{tabular}
	\end{center}
	\caption{\label{fig8} Depiction of the different forces for the energy density model-1 with $r_0 = 0.75 km$, $\alpha = -0.8 km$, $S_0 = 0.01 km^{-1}$, and $\lambda = 1.45km^{-1}$ (Left), for the energy density model-2 with $r_0 = 0.75 km$, $\alpha = -0.8 km$, $S_0 = 0.01 km^{-1}$, $\lambda = 1.45km^{-1}$, $\mu = 1 km^{-1}$, and $\nu = 1$ (Middle), and for the energy density model-3 with $r_0 = 0.75 km$, $\alpha = -0.8 km$, $S_0 = 0.01 km^{-1}$, $\lambda = 1.45km^{-1}$, $\nu$ =0.4, and $\beta = 1km^{-1}$ (Right).}
\end{figure}

The energy conditions are shown in Fig.~\ref{fig6}, illustrating that the 
Yukawa-Casimir energy density (\ref{rho3}) remains negative throughout the entire spacetime. Similar to the aforementioned models, the present solutions violate the NEC, DEC, and SEC near the wormhole throat. Therefore, the reported wormholes contain the exotic matter that yields the fuel to sustain them in the context of EC gravity.

\section{Matching of interior spacetime to exterior spacetime}\label{sec7}

For the wormhole to be asymptotically flat, it is required that $\Omega(r)/r \to 0$ as $r \to \infty$ and simultaneously $e^{2\Phi(r)} \to 1$ as $r \to \infty$. In the case of the chosen redshift function $\Phi(r) = \alpha/r$,  we indeed obtain $e^{2\Phi(r)} \to 1$ as $r \to \infty$. However, the present wormhole solutions reveal that $\Omega(r)/r \nrightarrow 0$ as $r \to \infty$, as is evident from Figs.~\ref{fig1}, \ref{fig3}, and \ref{fig5}. In particular, we have the following results 
 \begin{itemize}
     \item For energy density model-1 with $r_0 = 0.75 km$, $\alpha = -0.8 km$, $S_0 = 0.01 km^{-1}$, $\lambda = 1.45km^{-1}$, and $\Lambda = 0.01km^{-2}$:
     
       $\frac{\Omega(r)}{r} = \frac{r \left[31.84 \text{Ei}(-1.45 r)-0.012 \text{Ei}\left(\frac{1.6}{r}\right)+2.4 r^3-0.006 e^{\frac{1.6}{r}} \{r(r+0.8)+1.28\} r+535.188\right]+21.96 e^{-1.45 r}}{720 r^2}.$  
     
     \item For energy density model-2 with $r_0 = 0.75 km$, $\alpha = -0.8 km$, $S_0 = 0.01 km^{-1}$, $\lambda = 1.45km^{-1}$, $\mu = 1 km^{-1}$, $\nu = 1$, and $\Lambda = 0.01km^{-2}$:
     
     $\frac{\Omega(r)}{r} =\frac{ r \left[0.049 r \text{Ei}\left(\frac{1.6}{r}\right)-46.17 r \text{Ei}(-1.45 r)+r \left\{9.6 r^3-e^{\frac{1.6}{r}} \{(0.02 r+0.019)r+0.031\} r+2122.12\right\}+19.74\right]-(31.84r-21.96)e^{-1.45 r}}{2880 r^3}$.

     \item For energy density model-3 with $r_0 = 0.75 km$, $\alpha = -0.8 km$, $S_0 = 0.01 km^{-1}$, $\lambda = 1.45km^{-1}$, $\nu$ =0.4, $\beta = 1km^{-1}$, and $\Lambda = 0.01km^{-2}$:
     
     $\frac{\Omega(r)}{r} = \frac{r^2 \left[0.02 \text{Ei}\left(\frac{1.6}{r}\right)+\xi_{18}-e^{\frac{1.6}{r}} r \{(0.012 r+0.01) r+0.015\}\right]+e^{-1.45 r} (13.18\, -19.11 r)+e^{-r} (6.27 r-6.27)}{1440 r^3}$.
 \end{itemize}
Here, $\xi_{18}=4.8 r^3-27.7 \text{Ei}(-1.45r)+6.3 \text{Ei}(- r)+1076.9$. Note that $\text{Ei}(r)$ denotes the exponential integral function.  From the above results, it is evident that $\Omega(r)/r \nrightarrow 0$ as $r \to \infty$. Hence, the present wormhole solutions are non-asymptotically flat. To address this, the wormhole configurations are truncated at a finite radial distance $r = \mathcal{R}_1$ and smoothly matched to the exterior Schwarzschild spacetime, which is given by
\begin{equation}
ds^2 =- \left(1-\frac{2M_w}{r}\right)dt^2+\left(1-\frac{2M_w}{r}\right)^{-1}dr^2+r^2(d\theta^2 + \text{sin}^2\theta d\phi^2),\label{sch}
\end{equation}
where $M_w$ represents the mass of the wormhole. To ensure continuity at the junction surface, the wormhole metric (\ref{Metric}) is matched with the exterior Schwarzschild spacetime (\ref{sch}) at $r = \mathcal{R}_{1} > \mathcal{R}_{*}$, the Schwarzschild radius, yielding 
\begin{eqnarray}
 \Omega(\mathcal{R}_{1})=2M_w~~~ \text{and}~~~  \Phi(\mathcal{R}_1)=\frac{1}{2}\ln(1-2M_w/\mathcal{R}_1).\label{mc1}   
\end{eqnarray}


\begin{figure}[h]
	\begin{center}
		\begin{tabular}{rl}
			\includegraphics[width=5.6cm]{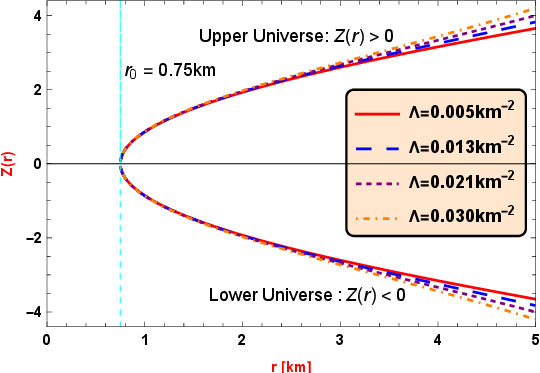}
			\includegraphics[width=5.6cm]{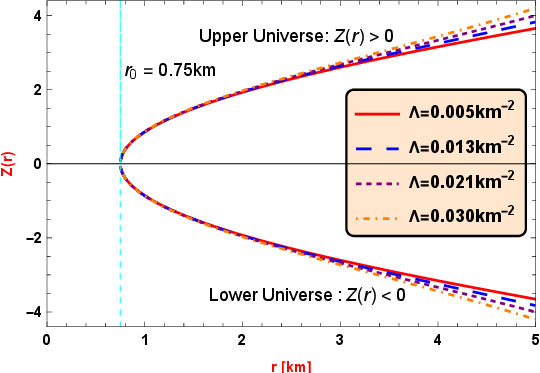}
			\includegraphics[width=5.6cm]{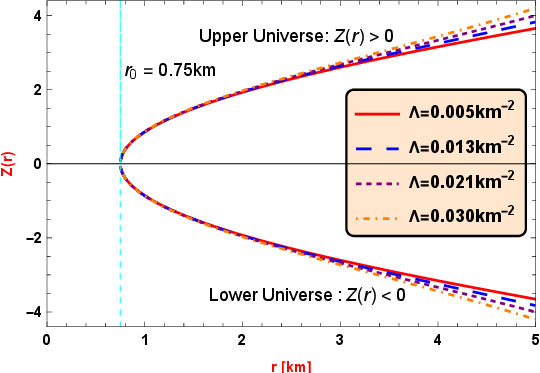}
			\\
		\end{tabular}
	\end{center}
	\caption{Depiction of the embedding surface for the energy density model-1 with $r_0 = 0.75 km$, $\alpha = -0.8 km$, $S_0 = 0.01 km^{-1}$, and $\lambda = 1.45km^{-1}$ (Left), for the energy density model-2 with $r_0 = 0.75 km$, $\alpha = -0.8 km$, $S_0 = 0.01 km^{-1}$, $\lambda = 1.45km^{-1}$, $\mu = 1 km^{-1}$, and $\nu = 1$ (Middle), and for the energy density model-3 with $r_0 = 0.75 km$, $\alpha = -0.8 km$, $S_0 = 0.01 km^{-1}$, $\lambda = 1.45km^{-1}$, $\nu$ =0.4, and $\beta = 1km^{-1}$ (Right). }\label{fig8a}
\end{figure}

Therefore, the metric describing the spacetime geometry of the constructed wormhole solutions can be expressed as 
\begin{eqnarray}
&&ds^2 =- e^{2\Phi(r)}dt^2+\left(1-\frac{\Omega(r)}{r}\right)^{-1}dr^2+r^2(d\theta^2 + \text{sin}^2\theta \,d\phi^2) \quad\text{for}\quad r<\mathcal{R}_1,
\\
\text{and}&&\nonumber
\\
&&ds^2 =-\left(1-\frac{2M_w}{r}\right)dt^2+\left(1-\frac{2M_w}{r}\right)^{-1}dr^2+r^2(d\theta^2 + \text{sin}^2\theta \,d\phi^2)
   \quad\text{for}\quad r\ge \mathcal{R}_1.
\end{eqnarray}

\section{Amount of exotic matter}\label{sec8}

Here, we examine the total quantity of exotic matter necessary to sustain the wormhole geometry, which entails a violation of the averaged null energy condition (ANEC). To quantify this, Visser et al.~\cite{mv03} proposed the volume integral quantifier ($\mathcal{VIQ}$), defined as
\begin{equation}
\mathcal{VIQ}=\oint\left[\rho(r)+P_r(r)\right]d\mathcal{V} = 8\pi\int_{r_{0}}^{\infty}r^2\left[\rho(r)+P_r(r)\right]dr.\label{i}
\end{equation}
In light of the present wormhole solutions, the $\mathcal{VIQ}$ can be expressed as
\begin{equation}
\mathcal{VIQ}= 8\pi\int_{r_{0}}^{\mathcal{R}_1} r^2\left[\rho(r)+\mathcal{P}_r(r)\right]dr.\label{I}
\end{equation}

Here, we want to examine the behavior of the volume integral quantifier ($\mathcal{VIQ}$) in the vicinity of the wormhole throat. Specifically, we aim to determine whether $\mathcal{VIQ} \rightarrow 0$ in the limit $\mathcal{R}_1 \rightarrow r_0^+$, where $r_0^+$ represents a value infinitesimally greater than the throat radius $r_0$. This condition is of particular importance, as it implies that the total amount of exotic matter required to sustain the wormhole structure can be made arbitrarily small, thereby ensuring that the wormhole solutions are physically more plausible within the given gravitational framework. It should be noted that, due to the complicated expression of $\rho(r)+\mathcal{P}_r(r)$ for each model, we will employ a numerical technique to estimate the $\mathcal{VIQ}$. In this context, we estimate the $\mathcal{VIQ}$ for each reported wormhole model in EC gravity from Eq. (\ref{I}) over $0.005km^{-2} \leq \Lambda \leq 0.03km^{-2}$, and the results are illustrated in Fig. \ref{fig7}. It is evident from Fig. \ref{fig7} that the $\mathcal{VIQ}$ remains negative near the throat for all wormhole models within the range $0.005km^{-2} \leq \Lambda \leq 0.03km^{-2}$. Furthermore, the magnitude of the negative $\mathcal{VIQ}$ increases as one moves away from the wormhole throat up to a certain radial distance, beyond which $\mathcal{VIQ}$  becomes positive. Notably, for all considered wormhole models, the negative values of $\mathcal{VIQ}$ approach zero in the vicinity of the throat within $0.005km^{-2} \leq \Lambda \leq 0.03km^{-2}$. 

To support the numerical claim, we estimate the $\mathcal{VIQ}$ analytically in the limit $R_1\to r_0^+$. For this, we consider $Q(r) = \rho(r)+\mathcal{P}_r(r)$, and set $r=r_0+\varepsilon$ with small $\varepsilon > 0$. Therefore, the Taylor series expansion yields the following result
\begin{eqnarray}
    Q(r) = Q(r_0+\varepsilon) = Q(r_0)+\varepsilon Q'(r_0)+\varepsilon^2 \frac{Q''(r_0)}{2!}+\varepsilon^3 \frac{Q'''(r_0)}{3!}+......
\end{eqnarray}
Now, the integrand of Eq. (\ref{I}) near the throat can be expressed as
\begin{eqnarray}
    r^2 Q(r) &=& (r_0+\varepsilon)^2\left(Q(r_0)+\varepsilon Q'(r_0)+\varepsilon^2 \frac{Q''(r_0)}{2!}+\varepsilon^3 \frac{Q'''(r_0)}{3!}+......\right)\\
    &=& r_0^2Q(r_0)+\left(2r_0Q(r_0)+r_0^2Q'(r_0)\right)\varepsilon+O(\varepsilon^2).
\end{eqnarray}

Thus, the $\mathcal{VIQ}$ can be expressed 
\begin{eqnarray}
\mathcal{VIQ}= 8\pi\int_{r_{0}}^{\mathcal{R}_1} r^2\left[\rho(r)+\mathcal{P}_r(r)\right]dr &=& 8\pi\int_{0}^{\mathcal{R}_1-r_0}\left[ r_0^2Q(r_0)+\left(2r_0Q(r_0)+r_0^2Q'(r_0)\right)\varepsilon+...\right]d\varepsilon,\nonumber
\\
&\simeq& 8\pi\left[r_0^2Q(r_0)(\mathcal{R}_1-r_0)+\frac{1}{2}\left(2r_0Q(r_0)+r_0^2Q'(r_0)\right)(\mathcal{R}_1-r_0)^2\right],\nonumber
\\
&\simeq& 8\pi\mathcal{R}_1 r_0(\mathcal{R}_1-r_0)Q(r_0).
\label{IFF}
\end{eqnarray}
It is found that $Q(r_0) = [\rho(r)+\mathcal{P}_r(r)]_{r=r_0} < 0$ for each of our models; therefore, the above result (\ref{IFF}) confirms that $\mathcal{VIQ}$ is negative near the wormhole throat. Moreover, $\mathcal{VIQ}$ tends to zero as $\mathcal{R}_1 \rightarrow r_0^+$, i.e., near the throat of the wormholes.
Therefore, the traversable wormholes within EC gravity can be supported by a minimal amount of exotic matter.

\section{Equilibrium Analysis}\label{sec9}

For any astrophysical object, achieving equilibrium is crucial to prevent collapse. Such a state arises from the balance among different acting forces. In the framework of Einstein-Cartan gravity, the equilibrium condition is described by the generalized Tolman-Oppenheimer-Volkoff (TOV) equation, expressed as

\begin{eqnarray}
\Phi'(r)\left[\rho(r)+P_r(r)\right]+P_r'(r)+\frac{2}{r}\left[P_r(r)-P_t(r)\right]-\frac{1}{2}\left[\Phi'(r)S^2+\frac{1}{2}(S^2)'\right]=0.\label{cq}
\end{eqnarray}
Since the spin contribution in the TOV equation is set to zero in Eq.~(\ref{sp}), 
the above TOV equation reduces to the following form 
\begin{eqnarray}
	&&-\Phi'(r)[\rho(r)+P_r(r)]-P'_r(r)+\frac{2}{r}[P_t(r)-P_r(r)]=0.\label{t}
\end{eqnarray}

We can write the above TOV equation (\ref{t}) in the following form 
\begin{eqnarray}
F_g(r)+F_h(r)+F_a(r)=0, 
\end{eqnarray}
where 
\begin{itemize}
    \item $ F_g(r) = -\Phi'(r)[\rho(r)+P_r(r)]$, termed as the gravitational force.
    \item $F_h(r) = -P'_r(r)$, termed  as the hydrostatic force.
    \item $F_a(r) = \frac{2}{r}[P_t(r)-P_r(r)]$, termed  as the anisotropic force.
\end{itemize}
Our reported wormhole solutions are found to remain in a stable equilibrium configuration due to the combined influence of the fundamental forces present in the system. In particular, the inward pull generated by the hydrostatic force $F_h(r)$  is effectively counterbalanced by the gravitational force $F_g(r)$ and the anisotropic force $F_a(r)$.  The simultaneous action and mutual adjustment of these forces prevent any instability, thereby ensuring the persistence of equilibrium, as depicted in Fig.~\ref{fig8}. 

\begin{figure}[h]
	\begin{center}
		\begin{tabular}{rl}
			\includegraphics[width=5.6cm]{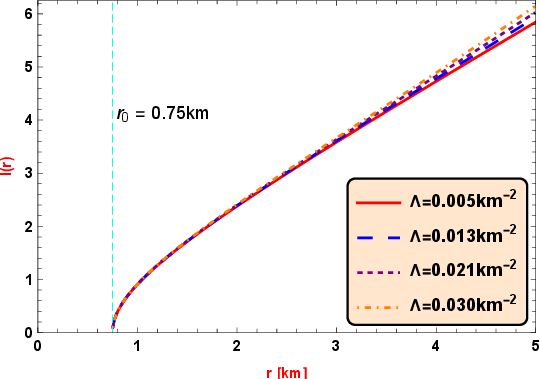}
			\includegraphics[width=5.6cm]{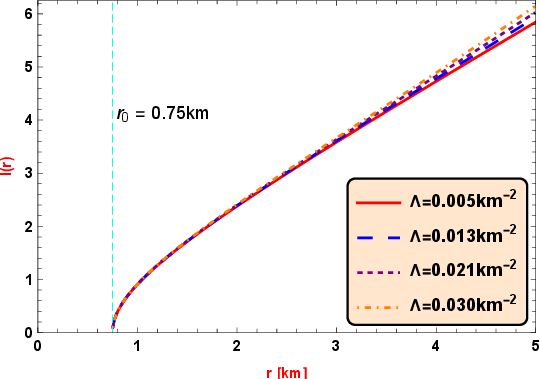}
			\includegraphics[width=5.6cm]{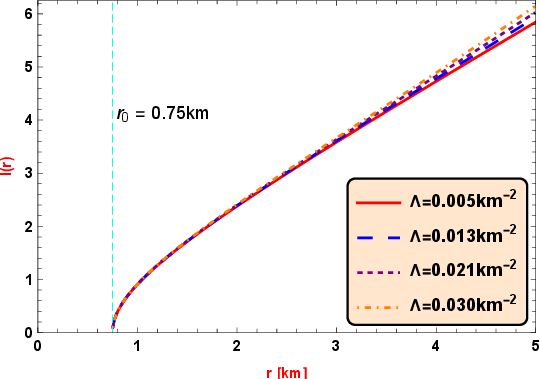}
			\\
		\end{tabular}
	\end{center}
	\caption{Depiction of the proper radial distance for the energy density model-1 with $r_0 = 0.75 km$, $\alpha = -0.8 km$, $S_0 = 0.01 km^{-1}$, and $\lambda = 1.45km^{-1}$ (Left), for the energy density model-2 with $r_0 = 0.75 km$, $\alpha = -0.8 km$, $S_0 = 0.01 km^{-1}$, $\lambda = 1.45km^{-1}$, $\mu = 1 km^{-1}$, and $\nu = 1$ (Middle), and for the energy density model-3 with $r_0 = 0.75 km$, $\alpha = -0.8 km$, $S_0 = 0.01 km^{-1}$, $\lambda = 1.45km^{-1}$, $\nu$ =0.4, and $\beta = 1km^{-1}$ (Right). }\label{fig8b}
\end{figure}

\section{Physical Features of Traversable Wormholes}\label{sec10}

In this section, we examine key features of the obtained wormhole solutions, including the embedding surface and proper radial distance, which provide insights into the geometric shape and visual representation of the wormhole structure, the tidal forces that determine traversability conditions, and the total gravitational energy that quantifies the overall energetic content of the configuration and reflects the strength of the underlying gravitational interaction.

\subsection{Embedding Surface and Proper Radial Distance}\label{sec10a}

The wormhole geometry is visualized through embedding diagrams by considering an equatorial slice $\theta=\pi/2$ at constant time $t$, which reduces the metric (\ref{Metric}) as
\begin{equation}
ds^2 = -\left(1-\frac{\Omega(r)}{r}\right)^{-1}dr^2-r^2d\phi^2.\label{es}
\end{equation}

In three-dimensional space, the embedded surface $Z(r)$ of the axially symmetric wormhole can be expressed as \cite{MT88}:

\begin{equation}
ds^2 = -\left[1+\left(\frac{d Z(r)}{dr}\right)^2\right]dr^2-r^2 d\phi^2.\label{es1}
\end{equation}

Thus, the results (\ref{es}) and (\ref{es1}) yield the following differential equation for the embedding surface $Z(r)$   
\begin{equation}
\frac{dZ(r)}{dr} = \pm\left(\frac{r}{\Omega(r)}-1\right)^{-\frac{1}{2}}.\label{dz1}
\end{equation}

\begin{figure}[h]
\begin{center}
\begin{tabular}{rl}
\includegraphics[width=5.8cm]{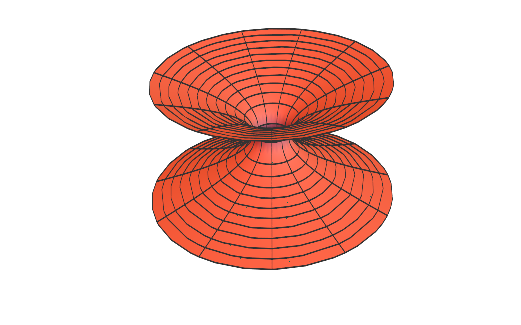}
\includegraphics[width=5.8cm]{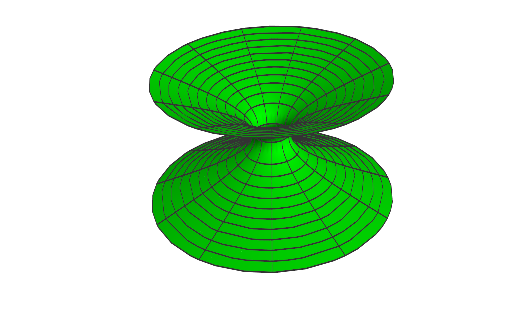}
\includegraphics[width=5.8cm]{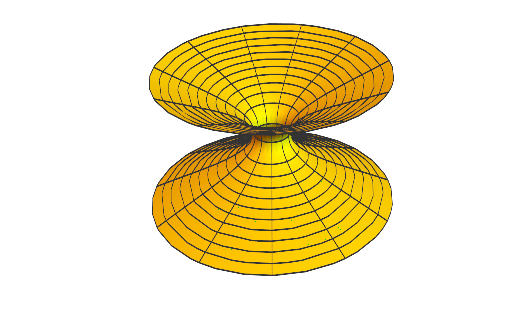}
\\
\end{tabular}
\end{center}
\caption{ Full visualization diagrams of wormholes for the energy density model-1 with $r_0 = 0.75 km$, $\alpha = -0.8 km$, $S_0 = 0.01 km^{-1}$, $\lambda = 1.45km^{-1}$, and $\Lambda = 0.013km^{-2}$  (Left), for the energy density model-2 with $r_0 = 0.75 km$, $\alpha = -0.8 km$, $S_0 = 0.01 km^{-1}$, $\lambda = 1.45km^{-1}$, $\mu = 1 km^{-1}$, $\nu = 1$, and $\Lambda = 0.013km^{-2}$  (Middle), and for the energy density model-3 with $r_0 = 0.75 km$, $\alpha = -0.8 km$, $S_0 = 0.01 km^{-1}$, $\lambda = 1.45km^{-1}$, $\nu$ =0.4, $\beta = 1km^{-1}$, and $\Lambda = 0.013km^{-2}$ (Right).}\label{fig8c}
\end{figure}

From Eq. (\ref{dz1}), it follows that $dZ(r)/dr$ diverges at the throat, making $Z(r)$ vertical there.  The embedding surface $Z(r)$ can nevertheless be expressed as
\begin{equation}
Z(r) = \pm \int_{r_0^+}^{r} \frac{dr}{\sqrt{r/\Omega(r)-1}},\label{z1}
\end{equation}
where the $\pm$ sign corresponds to the upper and lower universes of the wormhole geometry.

Furthermore, the proper radial distance is defined as \cite{MT88}
\begin{eqnarray}
  l(r) =  \int_{r_0^+}^{r} \frac{dr}{\sqrt{1-\Omega(r)/r}}.  
\end{eqnarray}

The embedding surfaces corresponding to the present wormhole solutions are shown in Fig.~\ref{fig8a}. These depict the spatial geometry of the wormhole by embedding a two-dimensional equatorial slice into three-dimensional Euclidean space. The region with positive curvature ($Z(r)>0$) represents the upper universe, while the region with negative curvature ($Z(r)<0$) corresponds to the lower universe, both joined through the wormhole throat. The proper radial distance of the present wormhole solutions exhibits a finite, monotonically increasing behavior, as shown in Fig.~\ref{fig8b}. Moreover, the full visualization diagrams of wormholes are depicted in Fig.~\ref{fig8c}. Together, these analyses provide deeper insight into the physical plausibility of the proposed wormhole models within the framework of EC gravity.

\subsection{Tidal Forces} \label{sec10b}

For a wormhole to be traversable, the tidal forces on a traveller must remain within human tolerance, not exceeding Earth’s surface gravity $g_\oplus\simeq 9.80 m/s^2$. Considering radial motion in the equatorial plane ($\theta=\pi/2$), the condition $|\Delta {\bf a}| \leq g_\oplus$ ensures traversability, leading to the following radial and lateral tidal forces constraints \cite{MT88}
\begin{eqnarray}
   \left|\left(1-\frac{\Omega(r)}{r}\right)\left(\Phi''(r)+\Phi'^2(r)-\frac{r\Omega'(r)-\Omega(r)}{2r(r-\Omega(r))}\Phi'(r)\right)c^2\right||\xi|&\leq& g_\oplus,\label{red}
    \\
 \left|\frac{\gamma^2}{2r^2}\left[\left(v/c\right)^2\left(\Omega'(r)-\Omega(r)/r\right)+2(r-\Omega(r))\Phi'(r)\right]\right||\xi|&\leq& g_\oplus,\label{lat}
\end{eqnarray}
where $\xi$ is the traveller's size, $v$ is the velocity, and $\gamma=\left(1-v^2/c^2\right)^{-1/2}$ is the Lorentz factor. For nonrelativistic motion, $v << c$ then $\gamma \approx 1$. Now, at the wormhole throat, the above tidal forces constraints for a traveller of body length $2m$ take the form
\begin{eqnarray}
    |\Phi'(r_0)|&\leq& \frac{g_\oplus r_0}{1-\Omega'(r_0)},\label{dphi}
    \\
    v &\leq& \frac{g_\oplus r_0}{\sqrt{1-\Omega'(r_0)}}.\label{v}
\end{eqnarray}

Therefore, the radial tidal condition \eqref{dphi} imposes a restriction on the redshift function, while the lateral condition \eqref{v} bounds the traveller’s velocity $v$. In particular, the radial constraint \eqref{dphi} further leads to the following restriction on the gravitational parameter $\Lambda$ 
\begin{eqnarray}
    \Lambda &\leq& \frac{\alpha  \left[180 r_0^3 S_0^2 e^{-2\alpha/r_0}+720r_0+\pi ^2\right]-720r_0^4 g_\oplus}{720 \alpha  d^3}~~\text{in the density model-1},
\\
  \Lambda &\leq&  \frac{\alpha\left[1440 r_0^2 + 360 r_0^4 S_0^2e^{-2\alpha/r_0}+\pi^2(\mu+\nu)\right] - 1440r_0^5 g_\oplus}{1440\alpha r_0^4}~~\text{in the density model-2},
  \\
  \Lambda &\leq&  \frac{\alpha \left[\pi^{2}(1-2\nu)- 720 r_0^{2}+180 r_0^{4}S_0^{2} e^{-2\alpha/r_0}\right] -720r_0^{5} g_\oplus}
{720 \alpha r_0^{4}}~~\text{in the density model-3}.
\end{eqnarray}

\begin{figure}[h]
\begin{center}
\begin{tabular}{rl}
\includegraphics[width=5.6cm]{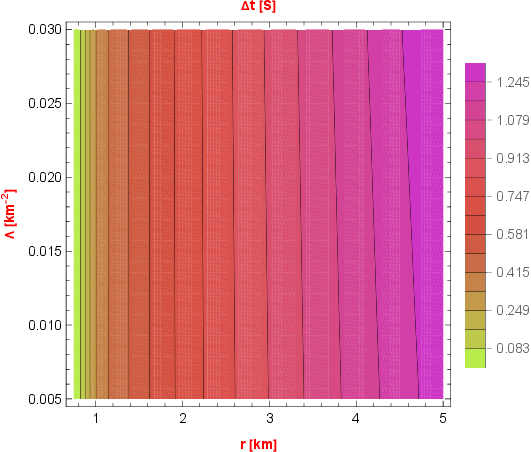}
\includegraphics[width=5.6cm]{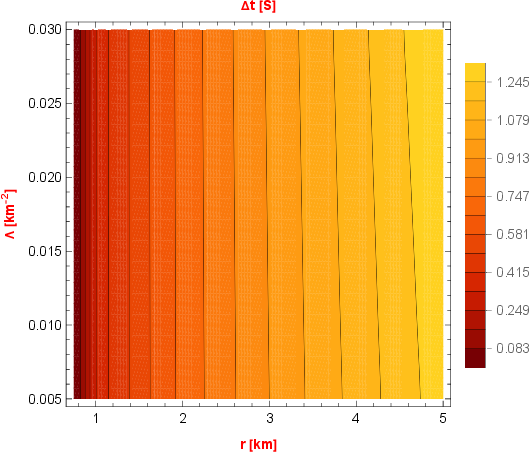}
\includegraphics[width=5.6cm]{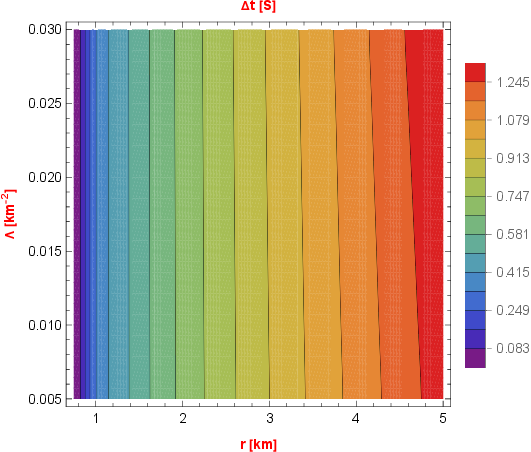}
\\
\end{tabular}
\end{center}
\caption{Depiction of the total time for the energy density model-1 with $r_0 = 0.75 km$, $\alpha = -0.8 km$, $S_0 = 0.01 km^{-1}$, and $\lambda = 1.45km^{-1}$ (Left), for the energy density model-2 with $r_0 = 0.75 km$, $\alpha = -0.8 km$, $S_0 = 0.01 km^{-1}$, $\lambda = 1.45km^{-1}$, $\mu = 1 km^{-1}$, and $\nu = 1$ (Middle), and for the energy density model-3 with $r_0 = 0.75 km$, $\alpha = -0.8 km$, $S_0 = 0.01 km^{-1}$, $\lambda = 1.45km^{-1}$, $\nu$ =0.4, and $\beta = 1km^{-1}$ (Right).}\label{fig9}
\end{figure}

Furthermore, condition (\ref{v}) yields the following constraints on the traveller’s velocity through the proposed wormholes 
\begin{eqnarray}
    v &\leq& \frac{r_0 g_\oplus}{\sqrt{1+\frac{r_0^2}{4} S_0^2 e^{-2\alpha/r_0}-\Lambda r_0^2+\frac{\pi ^2}{720r_0}}}~~\text{in the density model-1},
\\
  v &\leq&  \frac{r_0 g_\oplus}{\sqrt{1+\frac{r_0^2}{4} S_0^2 e^{-2\alpha/r_0}-\Lambda r_0^2+\frac{\pi ^2 (\mu +\nu )}{1440 d^2}}} ~~\text{in the density model-2}
  \\
  v &\leq&  \frac{r_0 g_\oplus}{\sqrt{1+\frac{r_0^2}{4} S_0^2 e^{-2\alpha/r_0}-\Lambda r_0^2+\frac{\pi ^2 (1-2 \nu )}{720 d^2}}}~~\text{in the density model-3}.
\end{eqnarray}

In this context, the total time for a traveller moving with constant speed $v$ is defined by
\begin{eqnarray}
    \Delta t = \int_{r_0}^{r} \frac{e^{-\Phi(r)}dr}{v\sqrt{1-\Omega(r)/r}}.
\end{eqnarray}

The variation of the total travel time is illustrated in Fig.~\ref{fig9}, showing the expected result that longer radial distances correspond to greater travel times. Moreover, the parameter $\Lambda$ has no significant influence on the total time near the wormhole throat, while its effect becomes noticeable farther away from the throat.

\subsection{Total Gravitational Energy} \label{sec10c}
The proposed wormhole structures are supported by exotic matter, arising from the violation of the NEC. Therefore, it is important to explore the characteristics of the total gravitational energy in this setting. The total gravitational energy $E_g$ is defined as \cite{dl07, kk09}
\begin{eqnarray}
  E_g = Mc^2 -E_M,\label{eg}  
\end{eqnarray}
where $Mc^2$ stands for the total energy, defined as
\begin{eqnarray}
    Mc^2 = \frac{1}{2}\int_{r_0}^rT_t^tr^2dr+\frac{r_0}{2}.\label{mc}
\end{eqnarray}

Here, the term $\frac{r_0}{2}$ in Eq. (\ref{mc}) corresponds to the effective mass \cite{kk09}. The quantity $E_M$ represents the total contribution from additional energy components such as rest energy, internal energy, kinetic energy, and similar forms, given by
\begin{eqnarray}
    E_M = \frac{1}{2}\int_{r_0}^r \frac{r^2 T_t^t dr}{\sqrt{1-\Omega(r)/r}}.\label{em} 
\end{eqnarray}

Thus, the results (\ref{mc}) and (\ref{em}) together reform  the total gravitational energy (\ref{eg}) as
\begin{eqnarray}
  E_g = \frac{1}{2}\int_{r_0}^r\left[1-\frac{1}{\sqrt{1-\Omega(r)/r}}\right]T_t^tr^2dr+\frac{r_0}{2}.\label{eg1}  
\end{eqnarray}

\begin{figure}[h]
	\begin{center}
		\begin{tabular}{rl}
			\includegraphics[width=5.2cm]{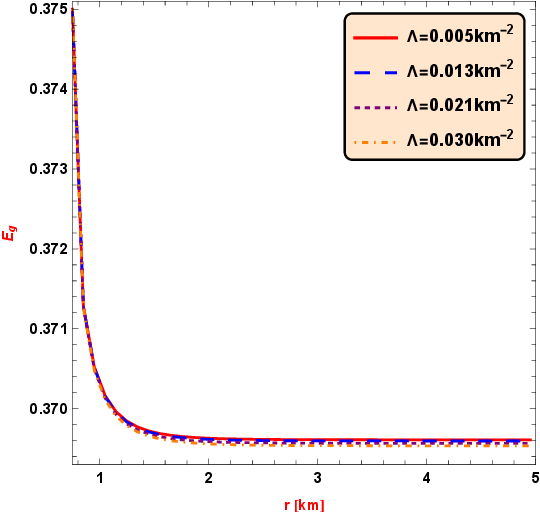}
			\includegraphics[width=5.2cm]{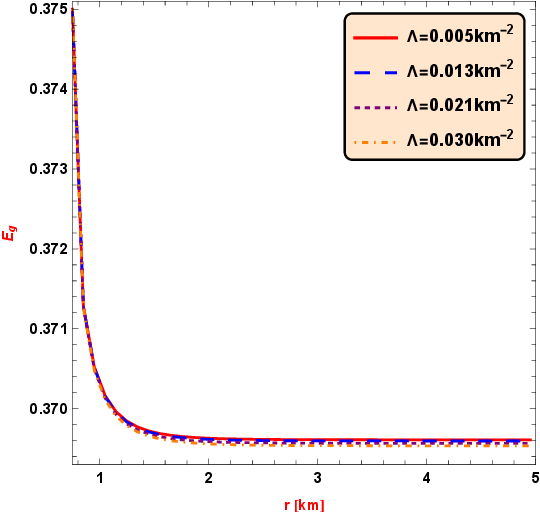}
			\includegraphics[width=5.2cm]{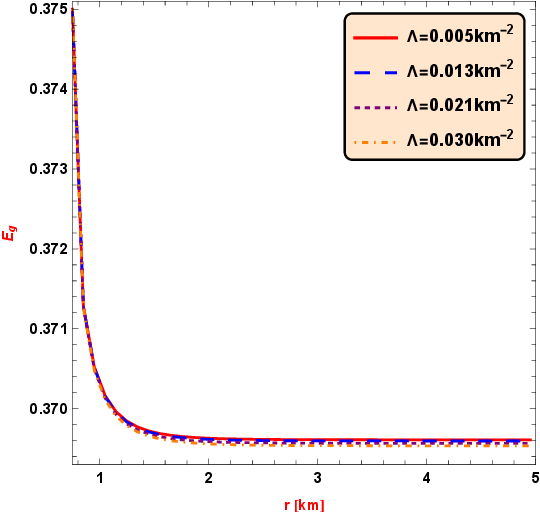}
			\\
		\end{tabular}
	\end{center}
	\caption{Depiction of the total gravitational energy for the energy density model-1 with $r_0 = 0.75 km$, $\alpha = -0.8 km$, $S_0 = 0.01 km^{-1}$, and $\lambda = 1.45km^{-1}$ (Left), for the energy density model-2 with $r_0 = 0.75 km$, $\alpha = -0.8 km$, $S_0 = 0.01 km^{-1}$, $\lambda = 1.45km^{-1}$, $\mu = 1 km^{-1}$, and $\nu = 1$ (Middle), and for the energy density model-3 with $r_0 = 0.75 km$, $\alpha = -0.8 km$, $S_0 = 0.01 km^{-1}$, $\lambda = 1.45km^{-1}$, $\nu$ =0.4, and $\beta = 1km^{-1}$ (Right).}\label{fig10}
\end{figure}

As noted by Misner et al. \cite{cw73}, the total gravitational energy is interpreted as attractive for $E_g < 0$, whereas it is regarded as repulsive for $E_g > 0$. Figure~\ref{fig10} illustrates the total gravitational energy $E_g$ for the present wormholes, showing that $E_g > 0$. This implies a repulsive gravitational energy, thereby supporting the possible existence of viable Yukawa-Casimir wormholes within the Einstein-Cartan gravity framework.

\section{Shadow of Wormholes}\label{sec11}

In this section, we analyze the shadows cast by the proposed wormhole solutions. 
To explore the influence of the wormhole on light propagation, we examine the trajectory of photons governed by the null geodesic equation in the static, spherically symmetric traversable wormhole spacetime, given by
\begin{equation}
ds^2 = -e^{2\Phi(r)}dt^2+\left(1-\frac{\Omega(r)}{r}\right)^{-1}dr^2+r^2(d\theta^2 + \sin^2\theta d\phi^2).\label{L}
\end{equation}

Here, we use the Hamilton–Jacobi equation to study the photon evolution in wormhole spacetime (\ref{L}), given by
\begin{eqnarray}
    \frac{\partial S}{\partial \sigma} = -\frac{1}{2} g^{ij} \frac{\partial S}{\partial x^i} \frac{\partial S}{\partial x^j},
\end{eqnarray}
where $\sigma$ is an affine parameter of the null geodesic. The Jacobi action $S$ can then be separated using the standard method as follows
\begin{eqnarray}
    S = \frac{1}{2} m^2\sigma- E t + L\phi + S_r(r) + S_\theta(\theta),
\end{eqnarray}
where $m$ is the particle's mass. In the case of a photon, we set $m = 0$. Moreover, $E$ and $L$ denote the photon's energy and angular momentum, respectively. Using the above equations, we obtain the following relations \cite{rs18}
\begin{eqnarray}
    \dot{t} &=& \frac{dt}{d\sigma}=\frac{E}{e^{2\Phi(r)}},
    \\
     \dot{r} &=& \frac{dr}{d\sigma}=\pm \sqrt{R(r)}\left(1-\Omega(r)/r\right)^{1/2}e^{-\Phi(r)},
     \\
     \dot{\theta}&=&\frac{d\theta}{d\sigma}=\pm \frac{1}{r^2}\sqrt{\Theta(\theta)},
    \\
    \dot{\phi} &=& \frac{d\phi}{d\tau}=\frac{L}{r^2\sin^2\theta}.
    \end{eqnarray}
with 
\begin{eqnarray}
    R(r) &=& E^2-K\frac{e^{2\Phi(r)}}{r^2},~~\Theta(\theta) = K -\frac{L}{\sin^2\theta},
\end{eqnarray}
where $K$ is the Carter constant.  For convenience in analyzing the geodesics, we define the dimensionless impact parameters as  $\xi = \frac{L}{E}$ and $\eta = \frac{K}{E^{2}}$.

\begin{figure}[h]
\begin{center}
\begin{tabular}{rl}
\includegraphics[width=6.2cm]{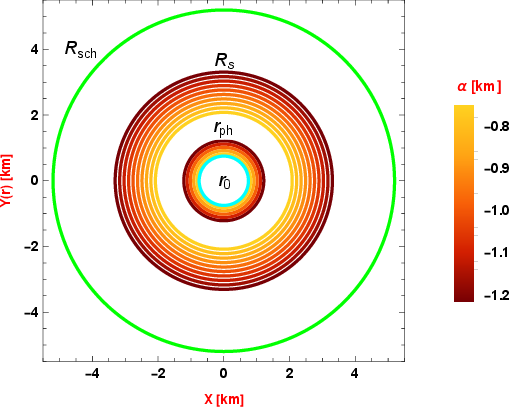}
\\
\end{tabular}
\end{center}
\caption{Depiction of the throat radius $r_0$, phothon sphere radius $r_{\rm ph}$, and shadow radius $R_s$ of the present wormholes.}\label{fig11}
\end{figure}

Now, a rescaling of the affine parameter $\sigma \to \sigma E$ yields the radial orbital equation of motion in terms of the effective potential $V_{\rm eff}(r)$ as follows
\begin{eqnarray}
    \left(\frac{dr}{d\sigma}\right)^2+V_{eff}=0,
\end{eqnarray}
where 
\begin{eqnarray}
    V_{eff} = -R(r)\left(1-\frac{\Omega(r)}{r}\right)e^{-2\Phi(r)} = -\left(1-\eta \frac{e^{2\Phi(r)}}{r^2}\right)\left(1-\frac{\Omega(r)}{r}\right)e^{-2\Phi(r)}.
\end{eqnarray}

Here, we focus on the spherical photon orbit of fixed radius $r = r_{\rm ph}$ with $\dot{r} = 0$ and $\ddot{r} = 0$. For a distant observer, photons approaching the wormhole (\ref{L}) near its equatorial plane follow the unstable circular orbits given by
\begin{eqnarray}
\mathcal{R}(r)|_{r=r_{\rm ph}} = 0,~ \mathcal{R}'(r)|_{r=r_{\rm ph}} = 0, ~ \text{and}~ R''(r)|_{r=r_{\rm ph}} \geq 0\label{RC}
\end{eqnarray}
The first condition yields the following result
\begin{eqnarray}
    \eta = \frac{r^2}{e^{2\Phi(r)}}\Big|_{r=r_{\rm ph}}.\label{eta}
\end{eqnarray}
Also, the second condition yields the photon sphere radius $r_{ph}$ as
\begin{eqnarray}
    \frac{d}{dr}\left(\frac{e^{2\Phi(r)}}{r^2}\right)\bigg|_{r=r_{\rm ph}} = 0 \quad\Rightarrow\quad
\left[2r\left(\Phi'(r)r -1\right)e^{2\Phi(r)}\right]_{r=r_{\rm ph}} = 0\quad\Rightarrow\quad r\Phi'(r)|_{r=r_{\rm ph}} = 1.\label{sc}
\end{eqnarray}
Therefore, for our wormhole solutions with the redshift function $\Phi(r) = \alpha/r$, the photon sphere radius $r_{\rm ph}$ is obtained from the above result (\ref{sc}) as
\begin{eqnarray}
  r_{\rm ph} = -\alpha.  
\end{eqnarray}
The above result indicates that the photon sphere radius exists only for $\alpha < 0$, more specifically, $\alpha \leq -r_0$. Notably, $R''(r)|_{r=r_{ph} = -\alpha}= \frac{2}{e^2\alpha^4} >0,$ indicating the unstable circular orbit. 

In the observer’s sky, the wormhole shadow can be described using the celestial coordinates $(X,Y)$, defined as \cite{jm72,se04,kh09}
\begin{eqnarray}
    X &=& \lim_{r_{\rm obs}\rightarrow \infty}\left(-r_{\rm obs}\sin\theta_0\left[\frac{d\phi}{dr}\right]_{r_{\rm obs},\theta_0}\right),\quad\quad
    Y = \lim_{r_{\rm obs}\rightarrow \infty}\left(r_{\rm obs}\left[\frac{d\theta}{dr}\right]_{r_{\rm obs},\theta_0}\right),\label{y}
\end{eqnarray}
where $\theta_{0}$ represents the inclination angle between the wormhole and the observer. Therefore, the celestial coordinates for the wormhole geometry can be expressed as \cite{pg13}
\begin{eqnarray}
    X &=& -\frac{\xi}{\sin\theta_0},\quad\quad Y = \pm \sqrt{\eta -\frac{\xi^2}{ \sin^2\theta_0}}.\label{y1}
\end{eqnarray}

For a static observer located at infinity, the radius of the wormhole shadow $R_{s}$, as viewed from the equatorial plane ($\theta_{0} = \pi/2$), is given by
\begin{eqnarray}
    R_{s} = \sqrt{X^2 + Y^2} = \sqrt{\eta}= \frac{r_{\rm ph}}{e^{\Phi(r_{\rm ph})}}=-\alpha e.\label{rs}
\end{eqnarray}

The above result and the boundary condition (\ref{mc1}) ensure that the existence of the shadow radius for the present wormhole solutions requires
\begin{eqnarray}
    \frac{\mathcal{R}_1}{2}\ln\left(1-2M_w/\mathcal{R}_1\right) < \alpha \leq -r_0.\label{alpha}
\end{eqnarray}
For $\alpha = \frac{\mathcal{R}_1}{2}\ln\left(1-2M_w/\mathcal{R}_1\right)$, the shadow is entirely determined by the external Schwarzschild solution, yielding a radius of $R_{Sch}=3\sqrt{3}M_w$. It is important to note that the photon sphere radius $r_{\rm ph}$ also satisfies the constraint (\ref{alpha}) on $\alpha$. We illustrate the throat radius, photon sphere radius, and shadow radius of the present wormhole solutions in Fig. \ref{fig11} corresponding to $-1.2165km <\alpha \leq -0.75km$ with $r_0 = 0.75 km$, $M_w = 1 M_\odot$, and $\mathcal{R}_1 = 6km$, showing that it increases as the parameter $\alpha$ decreases, i.e. the shadow radius is inversely related to the parameter $\alpha$.

\begin{figure}[h]
	\begin{center}
		\begin{tabular}{rl}
			\includegraphics[width=5.6cm]{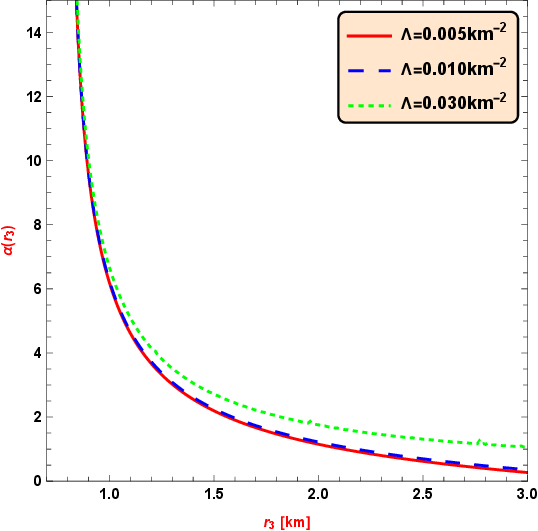}
			\includegraphics[width=5.6cm]{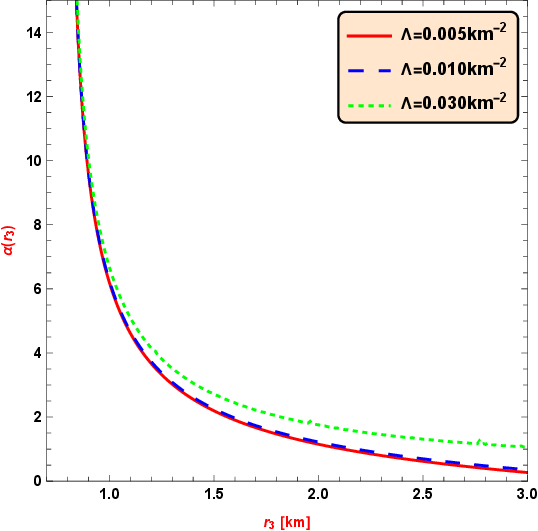}
			\includegraphics[width=5.6cm]{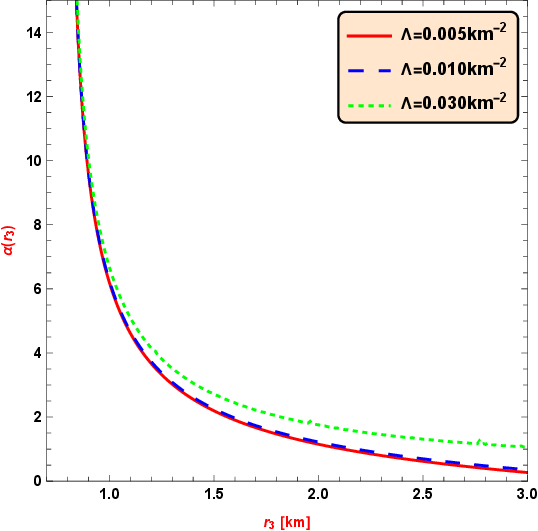}
			\\
		\end{tabular}
	\end{center}
	\caption{Depiction of the deflection angle for the energy density model-1 with $r_0 = 0.75 km$, $\alpha = -0.8 km$, $S_0 = 0.01 km^{-1}$, and $\lambda = 1.45km^{-1}$ (Left), for the energy density model-2 with $r_0 = 0.75 km$, $\alpha = -0.8 km$, $S_0 = 0.01 km^{-1}$, $\lambda = 1.45km^{-1}$, $\mu = 1 km^{-1}$, and $\nu = 1$ (Middle), and for the energy density model-3 with $r_0 = 0.75 km$, $\alpha = -0.8 km$, $S_0 = 0.01 km^{-1}$, $\lambda = 1.45km^{-1}$, $\nu$ =0.4, and $\beta = 1km^{-1}$ (Right).}\label{fig12}
\end{figure}

\section{Strong Deflection Angle}\label{sec12}

In this section, we investigate the influence of reported wormholes on the trajectory of light by evaluating the strong deflection angle. The seminal work of Virbhadra and collaborators \cite{ks98, cm01} demonstrated that strong gravitational lensing serves as a powerful probe of space-time geometry and astrophysical objects. Building on this, Bozza \cite{vb02} developed a systematic analytical method to study light deflection in any spherically symmetric space-time under strong gravitational fields. For a spherically symmetric spacetime with the line element
\begin{eqnarray}
    ds^2 = -A(r)dt^2+B(r)dr^2+C(r)(d\theta^2+\sin^2\theta d\phi^2),
\end{eqnarray}
the deflection angle $\alpha(r_c)$ in terms of the closest approach $r_c$ can be expressed as \cite{vb01, vb02}
\begin{equation}
\alpha(r_c) = -\pi + 2 \int_{r_c}^{\infty} \frac{\sqrt{B(r)}dr}{\sqrt{C(r)} \sqrt{\frac{C(r)}{C(r_c)}\frac{A(r_c)}{A(r)} - 1}}.
\end{equation}
Now, we can express the integrand as
\begin{eqnarray}
    I = \frac{\sqrt{B(r)}dr}{\sqrt{C(r)} \sqrt{\frac{C(r)}{C(r_c)}\frac{A(r_c)}{A(r)} - 1}}=\frac{\sqrt{B(r)}dr}{\sqrt{C(r)}\sqrt{\frac{C(r)}{A(r)}} \sqrt{\frac{A(r_c)}{C(r_c)} - \frac{A(r)}{C(r)}}}=\frac{\sqrt{A(r)}dr}{C(r)\sqrt{\frac{1}{B(r)}\left(\frac{A(r_c)}{C(r_c)} - \frac{A(r)}{C(r)}\right)}}.
\end{eqnarray}
Thus, the deflection angle for the present wormhole (\ref{Metric}) can be written as
\begin{eqnarray}
\alpha(r_c)=-\pi+2\int_{r_{c}}^{\infty}\frac{e^{\Phi(r)}dr}{r^2\sqrt{\left(1-\frac{\Omega(r)}{r}\right)\left(\frac{1}{b^{2}}-\frac{e^{2\Phi(r)}}{r^{2}}\right)}}, \label{DA}
\end{eqnarray}
where $b$ is the impact parameter, given by
\begin{eqnarray}
b=r_c e^{-\Phi(r_c)}. \label{beta}
\end{eqnarray}

The obtained deflection angle as a function of the closest approach distance $r_c$ is illustrated in Fig.~\ref{fig12}. The plot shows that the deflection angle decreases as the closest approach distance $r_c$ increases, indicating weaker bending farther from the wormhole. In contrast, as the light nears the throat, the deflection angle rises sharply and diverges, reflecting the intense curvature and strong gravitational effects in that region. Moreover, the parameter $\Lambda$ significantly influences the deflection angle in all wormhole configurations. Larger values of $\Lambda$ result in stronger light bending, indicating that $\Lambda$ directly enhances the spacetime curvature and gravitational lensing effects of the wormhole.

\section{Conclusion}\label{sec13}
The Einstein-Cartan (EC) gravity theory is a natural extension of General Relativity (GR) formulated within the Riemann-Cartan geometry that incorporates torsion \cite{EA24, EA25, FW07}. It can be regarded as a specific limit of the Poincaré gauge theory of gravity \cite{fw76, vn85, av06}. Within this framework, EC gravity provides a promising foundation for supporting and sustaining traversable wormhole configurations \cite{MR17, MR17A, ns24, MR19, KA16}. In this study, we have investigated the Yukawa-Casimir wormhole solutions sustained by the Yukawa-Casimir energy density within the framework of EC gravity. By introducing three distinct Yukawa-Casimir energy density profiles into the EC gravity framework, we derive three new shape functions and conduct a detailed analysis of their properties. The shape functions $\Omega(r)$ obtained from the three Yukawa-Casimir energy density models increase monotonically with the radial coordinate $r$ and satisfies the key traversable wormhole requirements: $\Omega(r)/r \leq 1$ for $r \geq r_0$ and the flaring-out condition $\Omega'(r) < 1$ for $r \geq r_0$ (see Figs. \ref{fig1}, \ref{fig3}, and \ref{fig5}). These results validate the suitability of the obtained shape functions for constructing traversable wormhole configurations in the EC gravity framework under appropriate parameter choices. However, the derived shape function fails to satisfy the asymptotic flatness condition, necessitating that the wormhole geometry be truncated at a finite radial distance and smoothly matched to an exterior Schwarzschild solution.  Nevertheless, the Yukawa-Casimir energy density profiles driven shape functions offer important insights into traversable wormhole geometries within the EC gravity framework, establishing a meaningful connection between astrophysical phenomena and the modified theory of gravity. The stability and persistence of a wormhole strongly depend on the nature of its matter content. To explore this aspect, we have analyzed the standard energy conditions, namely, NEC, DEC, and SEC, within the framework of the proposed wormhole geometries. For the proposed wormhole models, detailed graphical illustrations of the energy conditions are provided in Figs. \ref{fig2}, \ref{fig4}, and \ref{fig6}. These figures clearly demonstrate that, for all the proposed wormhole configurations, the NEC,  DEC, and SEC are violated in the vicinity of the wormhole throat under the same choices of appropriate parameters. This violation confirms the inevitable presence of exotic matter, which plays a crucial role in sustaining traversable wormhole geometries within the framework of EC gravity. In this context, we have analyzed the total amount of exotic matter using the volume integral quantifier ($\mathcal{VIQ}$). As illustrated in Fig. \ref{fig7}, the $\mathcal{VIQ}$ stays negative near the wormhole throat for all models within the range $0.005km^{-2} \leq \Lambda \leq 0.03km^{-2}$, and tends to approach zero in the immediate vicinity of the throat. Therefore, the behavior of $\mathcal{VIQ}$ indicates that traversable wormholes in EC gravity can be sustained with only a minimal amount of exotic matter for $0.005km^{-2} \leq \Lambda \leq 0.03km^{-2}$, within which the NEC is violated for all the proposed wormhole models. The stability of the obtained wormhole solutions arises from the balance among the fundamental forces of the system. The inward pull generated by the hydrostatic force is precisely counterbalanced by the gravitational force and the anisotropic force. This force equilibrium suppresses instabilities and guarantees the sustained stability of the configuration, as illustrated in Fig.~\ref{fig8}. To gain deeper insights into the present wormhole solutions, we have undertaken a detailed investigation of their key physical characteristics. In particular, we have examined the embedding surface, which provides a geometrical visualization of the wormhole structure, and the proper radial distance, which characterizes the spatial geometry and accessibility of the throat. Additionally, we have analyzed the tidal forces experienced by a hypothetical traveller, ensuring the traversability conditions of the wormhole, as well as the total gravitational energy, which offers valuable information about the overall energetic stability of the configuration. The embedding surface shown in Fig.~\ref{fig8a} illustrates the wormhole geometry by mapping a two-dimensional equatorial slice into three-dimensional Euclidean space. In this representation, the regions $Z(r) > 0$ and $Z(r)< 0$ correspond to the upper and lower universes, smoothly connected at the throat, thereby highlighting the bridge-like symmetry of the wormhole. The proper radial distance of the present wormhole solutions is finite and increases monotonically with the radial coordinate, as shown in Fig.~\ref{fig8b}. This behavior highlights the traversable nature of the wormhole geometries. In addition, Fig.~\ref{fig8c} presents the complete visualization diagrams of the wormhole geometries, providing a clear representation of their overall shape and spatial structure.  The analysis of tidal forces imposes specific restrictions on the gravitational parameter $\Lambda$ and places constraints on the velocity of a traveller passing through the proposed wormholes. In this regard, we have also analyzed the total time for a traveller moving with constant speed in Fig.~\ref{fig9},  showing the expected outcome that larger radial distances correspond to longer traversal times. The total gravitational energy associated with our wormhole solutions is found to be repulsive in nature (see Fig.~\ref{fig10}). This repulsive character further supports the potential existence of wormholes sustained by Yukawa–Casimir energy density profiles within the framework of EC gravity.  We have also analyzed the photon sphere radius and shadow radius of the wormholes, which are found to increases as the parameter $\alpha$ decreases, indicating an inverse relationship of the photon sphere radius and shadow radius with the parameter $\alpha$ (see Fig. \ref{fig11}).  Finally, using Bozza’s method, we have determined the strong deflection angle in the proposed wormhole spacetimes. The deflection angle decreases with increasing $r_c$ but diverges near the throat. Moreover, the larger values of the EC gravity parameter $\Lambda$ enhance the lensing effect by strengthening the spacetime curvature (see Fig.~\ref{fig12}).

Overall, our analyses demonstrate that the reported wormhole solutions sustained by Yukawa-Casimir energy density profiles are physically consistent within the EC gravity framework. The viability of these configurations fundamentally relies on the inevitable presence of exotic matter in all models considered. Thus, EC gravity provides a robust and fertile platform for realizing equilibrium stable Yukawa-Casimir traversable wormholes, thereby advancing the theoretical exploration of traversable wormhole geometries under modified gravity. Looking ahead, this study may motivate further investigations into broader classes of Yukawa-Casimir wormhole solutions within other modified theories of gravity.

 \section*{Acknowledgement}
We sincerely thank the anonymous reviewer(s) for their insightful comments and constructive suggestions, which have greatly enhanced the clarity and overall quality of this work.


\begin{thebibliography}{37}

\bibitem{WHM1} M.S. Morris, K.S. Thorne, {\it Am. J. Phys.} { \bf 56}, 395 (1988).

\bibitem{WHM2} M.S. Morris, K.S. Thorne, U. Yurtsever, {\it Phys. Rev. Lett.} { \bf 61}, 1446 (1988).

\bibitem{WHM3} S. Kar, N. Dadhich, M. Visser, {\it Pramana J. Phys.} {\bf 63}, 859 (2004).

\bibitem{WHM4} D. Hochberg, M. Visser, {\it Phys. Rev. D}  {\bf 56}, 4745 (1997).

\bibitem{WHM5} M. Visser, S. Kar, N. Dadhich, {\it Phys. Rev. Lett.}  {\bf 90}, 201102 (2003).

\bibitem{WHM6} E. Poisson, M. Visser, {\it Phys. Rev. D} { \bf 52}, 7318 (1995).

\bibitem{WHM7} S. W. Kim,  Phys. Lett.   {\bf A 166}, 13 (1992).

\bibitem{WHM8} F.S.N. Lobo, {\it Class. Quantum Gravity}  {\bf 21}, 4811 (2004).

\bibitem{WHM9} J.P.S. Lemos, F.S.N. Lobo, {\it Phys. Rev. D}  {\bf 69}, 104007 (2004).

\bibitem{WHM10} E.F. Eiroa, C. Simeone, {\it Phys. Rev. D}  {\bf 70}, 044008 (2004).

\bibitem{WHM11} E.F. Eiroa, C. Simeone, {\it Phys. Rev. D}  {\bf 71}, 127501 (2005).

\bibitem{WHM12} F. Rahaman, M. Kalam, S. Chakraborty, {\it Gen. Relativ. Gravit.}  {\bf 38}, 1687 (2006).

\bibitem{WHM13} C. Bejarano, E.F. Eiroa, C. Simeone, {\it Phys. Rev. D}  {\bf 75}, 027501 (2007).

\bibitem{WHM14} M. Thibeault, C. Simeone, E.F. Eiroa, {\it Gen. Relativ. Gravit.}  {\bf 38}, 1593 (2006).

\bibitem{WHM15} F. Rahaman, M. Kalam, S. Chakraborty, {\it Int. J. Mod. Phys. D}  {\bf 16}, 1669 (2007).

\bibitem{WHM16} E. Gravanis, S. Willison, {\it Phys. Rev. D} { \bf 75}, 084025 (2007).

\bibitem{WHM17} A.G. Agnese, M. La Camera, {\it Phys. Rev. D} { \bf 51}, 2011 (1995).

\bibitem{WHM18} K.K. Nandi, A. Islam, J. Evans, {\it Phys. Rev. D} { \bf 55}, 2497 (1997).

\bibitem{WHM19} F.S.N. Lobo, M.A. Oliveira, {\it Phys. Rev. D} { \bf 81}, 067501 (2010).

\bibitem{WHM20} S.V. Sushkov, S.M. Kozyrev, {\it Phys. Rev. D} { \bf 84}, 124026 (2011).

\bibitem{WHM21} F.S.N. Lobo, M.A. Oliveira, {\it Phys. Rev. D} { \bf 80}, 104012 (2009).

\bibitem{WHM22} N.M. Garcia, F.S.N. Lobo, {\it Phys. Rev. D} { \bf 82}, 104018 (2010).

\bibitem{WHM23} N. Montelongo Garcia, F.S.N. Lobo, {\it Class. Quantum Gravity} { \bf 28}, 085018 (2011).

\bibitem{WHM24} E.F. Eiroa, G.F. Aguirre, {\it Eur. Phys. J. C} { \bf 72}, 2240 (2012).

\bibitem{WHM25} M. Richarte, C. Simeone, {\it Phys. Rev. D} { \bf 80}, 104033 (2009).

\bibitem{WHM26} M.K. Zangeneh, F.S.N. Lobo, M.H. Dehghani, {\it Phys. Rev. D} { \bf 92}, 124049 (2015).

\bibitem{WHM27} V.D. Dzhunushaliev, D. Singleton, {\it Phys. Rev. D} { \bf 59}, 064018 (1999).

\bibitem{WHM28} J.P. de Leon, {\it J. Cosmol. Astropart. Phys.} { \bf 11}, 013 (2009).

\bibitem{WHM29} R. Shaikh, S. Kar, {\it Phys. Rev. D} { \bf 94}, 024011 (2016).

\bibitem{WHM30} A. Anabalon, A. Cisterna, {\it Phys. Rev. D} { \bf 85}, 084035 (2012).

\bibitem{WHM31} J.P.S. Lemos, F.S.N. Lobo, S.Q. Oliveira, {\it Phys. Rev. D} { \bf 68}, 064004 (2003).

\bibitem{WHM32} E. Elizalde, M. Khurshudyan, {\it Phys. Rev. D} { \bf 98}, 123525 (2018).

\bibitem{WHM33} P.H.R.S. Moraes, P.K. Sahoo, {\it Eur. Phys. J. D} { \bf 79}, 677 (2019).

\bibitem{WHM34} P.H.R.S. Moraes, P.K. Sahoo, {\it Phys. Rev. D} { \bf 96}, 044038 (2017).

\bibitem{WHM35} P.H.R.S. Moraes, P. Sahoo, {\it Eur. Phys. J. C} { \bf  78}, 46 (2018).

\bibitem{WHM36} P.K. Sahoo, P.H.R.S. Moraes, P. Sahoo, G. Ribeiro, {\it Int. J. Mod. Phys. D} { \bf 27}, 1950004 (2018).

\bibitem{WHM37} E. Elizalde, M. Khurshudyan, {\it Phys. Rev. D} { \bf 99}(2), 024051 (2019).

\bibitem{WHM38} P.H.R.S. Moraes, W. de Paula, R.A.C. Correa, {\it Int. J. Mod. Phys. D} { \bf 28}, 1950098 (2019).

\bibitem{WHM39} S. Kar, {\it Phys. Rev. D} { \bf 49}, 862 (1994).

\bibitem{WHM40} S. Kar, D. Sahdev, {\it Phys. Rev. D} { \bf 53}, 722 (1996).

\bibitem{WHM41} H. Maeda, T. Harada, B.J. Carr, {\it Phys. Rev. D} { \bf 79}, 044034 (2009).

\bibitem{WHM42} A.V.B. Arellano, F.S.N. Lobo, {\it Class. Quantum Gravity} { \bf 23}, 5811 (2006).

\bibitem{WHM43} M. Cataldo, P. Meza, P. Minning, {\it Phys. Rev. D} { \bf 83}, 044050 (2011).

\bibitem{CE1} H. Casimir, {\it Proc. Kon. Ned. Akad. Wetenschap} { \bf 51}, 793 (1948).

\bibitem{CE2} M. Sparnaay, {\it Nature} { \bf 180}, 334 (1957).

\bibitem{CE3} U. Mohideen, A. Roy, {\it Phys. Rev. Lett.} { \bf 81}, 4549 (1998).

\bibitem{CE4} G. Bressi, et al., {\it Phys. Rev. Lett.} { \bf 88}, 041804 (2002).

\bibitem{CE5} S. Vezzoli, et al., {\it Commun. Phys.} { \bf 2}, 84 (2019).

\bibitem{CE6} F. Wilczek, E.V. Linder, M.R.R. Good, {\it Phys. Rev. D} { \bf 101}, 025012 (2020).

\bibitem{CE7} R. Garattini, {\it Eur. Phys. J. C} { \bf 79}, 951 (2019).

\bibitem{CE8} R. Garattini, {\tt In Spacetime Physics and Gravitation}, World Scientific, (2022). 

\bibitem{CE9} R. Garattini, {\it Eur. Phys. J. C} { \bf 81}, 824 (2021).

\bibitem{CE10} A. Jawad, M.B. Amin Sulehri, S. Rani, {\it Eur. Phys. J. Plus} { \bf 137}, 1274 (2022).

\bibitem{CE11} A.K. Mishra, Shweta, U.K. Sharma, {\it Universe} { \bf 9}, 161 (2023).

\bibitem{CG1} E. Cartan, {\it Ann. Ec. Norm. Sup.} { \bf 40}, 325 (1923).

\bibitem{CG2} M.A. Lledó, L. Sommovigo, {\it Class. Quantum Gravity} { \bf 27}, 065014 (2010).

\bibitem{CG3} M. Mathisson, {\it Acta Phys. Pol.} { \bf 6}, 163 (1937).

\bibitem{CG4} H. Honl, A. Papapetrou, {\it Z. Phys.} { \bf 114}, 153 (1940).

\bibitem{CG5} D.D. Ivanenko, A.S. Gololobova, V.G. Krechet, V.G. Lapchinskii, Sov. Phys. J. { \bf 16}, 1680 (1973).

\bibitem{CG6} V.G. Krechet, M.L. Filchenkov, G.N. Shikin, {\it Gravit. Cosmol.} { \bf 14}, 292 (2008).

\bibitem{CG7} R.T. Jantzen, {\it J. Math. Phys.} { \bf 23}, 1137 (1982).

\bibitem{CG8} T. Watanabe, M.J. Hayashi, arXiv:gr-qc/0409029 (2004).

\bibitem{CG9} F. Hehl, P. von der Heyde, G.D. Kerlick, {\it Phys. Rev. D} { \bf  10}, 1066 (1974).

\bibitem{CG10} F.W. Hehl, P. von der Heyde, G.D. Kerlick, J.M. Nester, {\it Rev. Mod. Phys.} { \bf 48}, 393 (1976).

\bibitem{CG11} F.W. Hehl, {\it Found. Phys.} { \bf 15}, 451 (1985).

\bibitem{CG12} F.W. Hehl, J.D. McCrea, E.W. Mielke, Y. Ne’eman, {\it Phys. Rep.} { \bf 258}, 1 (1995).

\bibitem{CG13} A. Trautman, {\tt Symposia Mathematica}, Vol. 12 (Bologna) (1973).

\bibitem{CG14} A. Trautman, {\it Nature} { \bf 242}, 7 (1973).

\bibitem{CG15} W. Kopczyński, {\it Phys. Lett. A} { \bf 39}, 219 (1972).

\bibitem{CG16} W. Kopczyński, {\it Phys. Lett. A} { \bf 43}, 63 (1973).

\bibitem{CG17} J. Weyssenhoff, A. Raabe, {\it Acta Phys. Pol.} { \bf 9}, 7 (1947).

\bibitem{CG18} J. Weyssenhoff, {\tt Max-Planck-Festschrift}, Deutscher Verlag der Wissenschaften, Berlin, 155 (1958).

\bibitem{CG19} G. de Berredo-Peixoto, E.A. de Freitas, {\it Int. J. Mod. Phys. A} { \bf 24}, 1652 (2009).

\bibitem{CG20} S. D. Brechet, M. P. Hobson, A. N. Lasenby, {\it Class. Quantum Gravity} { \bf 24}, 6329 (2007).

\bibitem{EC1} S. Capozziello, S. Nojiri, S.D. Odintsov, {\it Phys. Lett. B} { \bf 781}, 99 (2018).  

\bibitem{EC2} K. Atazadeh, A. Khaleghi, H.R. Sepangi, Y. Tavakoli, {\it Int. J. Mod. Phys. D} { \bf 18}, 1101 (2009).  

\bibitem{EC3} D. Liu, M.J. Reboucas, {\it Phys. Rev. D} { \bf 86}, 083515 (2012).

\bibitem{EC4} M. Zubair, S. Waheed, {\it Astrophys. Space Sci.} { \bf 355}, 361 (2015).

\bibitem{EC5} T. Azizi, M. Gorjizadeh, {\it Europhys. Lett.} { \bf 117}, 60003 (2017). 

\bibitem{EC6} N.M. Garcia, et al., {\it Phys. Rev. D} { \bf 83}, 104032 (2011).  

\bibitem{EC7} K. Bamba, C.Q. Geng, S. Nojiri, S.D. Odintsov, {\it Phys. Rev. D} { \bf  79}, 083014 (2009). 

\bibitem{EC8} M. Sharif, A. Ikram, {\it Eur. Phys. J. C} { \bf 76}, 640 (2016). 

\bibitem{EC9} Z. Yousaf, et al., {\it Int. J. Geom. Methods Mod. Phys.} { \bf 15}, 1850146 (2018).  

\bibitem{EC10} S. Mandal, P.K. Sahoo, J. Santos, {\it Phys. Rev. D} { \bf 102}, 024057 (2020).  

\bibitem{EC11} S. Arora, J. Santos, P.K. Sahoo, {\it Phys. Dark Univ.} { \bf 31}, 100790 (2021). 

\bibitem{fw76} F. W. Hehl, P. Von der Heyde, and G. D. Kerlick and J. M. Nester, {\it Rev. Mod. Phys.} {\it 48}, 393 (1976).

\bibitem{vd86} V. De Sabbata and M. Gasperini, {\tt Introduction to Gravitation}, (World Scientific, Singapore, 1986).

\bibitem{vd90} V. De Sabbata and C. Sivaram, {\it Astrophysics and Space Science}, {\bf 165}, 51 (1990).

\bibitem{vd94} V. De Sabbata and C. Sivaram, {\tt 
Spin and Torsion in Gravitation}, (World Scientic, Singapore, 1994).

\bibitem{nj34} N. J. Poplawski, arXiv:0911.0334 [gr-qc].

\bibitem{tw61} T. W. B. Kibble, {\it J. Math. Phys.} {\bf 2}, 212 (1961).

\bibitem{dw62} D. W. Sciama, {\it Recent Developments in General Relativity}, p. 415 (Pergamon, 1962).

\bibitem{dw64} D. W. Sciama, {\it Rev. Mod. Phys.} {\bf 36}, 463 (1964).

\bibitem{dw64a} D. W. Sciama, {\it Rev. Mod. Phys.} {\bf 36}, 1103 (1964).

\bibitem{fw71} F. W. Hehl and B. K. Datta, {\it J. Math. Phys.} {\bf 12}, 1334 (1971).

\bibitem{fw71a} F. W. Hehl, {\it Phys. Lett. A} {\bf 36}, 225 (1971).

\bibitem{rt02} R. T. Hammond, {\it Rep. Prog. Phys.} {\bf 65}, 599 (2002).

\bibitem{dn16} D. N. Blaschke, F. Gieres, M. Reboud and M. Schweda, {\it Nucl. Phys. B} {\bf 912}, 192 (2016).

\bibitem{ea76} E. A. Lord, Tensor, Relativity and Cosmology, (McGraw-Hill, New Delhi, 1976).

\bibitem{fw73} F. W. Hehl, {\it Gen. Relativ. Gravit.} {\bf 4}, 333 (1973).

\bibitem{fw74} F. W. Hehl, {\it Gen. Relativ. Gravit.} {bf 5}, 491 (1974).

\bibitem{yn87} Y. N. Obukhov and V. A. Korotky, {\it Class. Quantum Grav.} {\bf 4}, 1633 (1987).

\bibitem{jw47} J. Weyssenhoff, A. Raabe, {\it Acta Phys. Pol.} {\bf 9}, 7 (1947).

\bibitem{jr83} J. R. Ray and L. L. Smalley, {\it Phys. Rev. D} {\bf 27}, 1383 (1983). 

\bibitem{ga74} G. A. Maugin, {\it Ann. Inst. Henri Poincaré} {\bf 20}, 41 (1974). 

 \bibitem{MG86} M. Gasperini, {\it Phys. Rev. Lett.} {\bf 56}, 2873 (1986).
 
\bibitem{MT88} M. S. Morris, K.S. Thorne, {\it Am. J. Phys.} {\bf 56}, 395 (1988).


\bibitem{ms88a} M. S. Morris, K. S. Thorne and U. Yurtsever, {\it Phys. Rev. Lett.} {\bf 61}, 1446 (1988).

\bibitem{ar55} A. Raychaudhuri, {\it Phys. Rev.} {\bf 98}, 1123 (1955).

 \bibitem{hy35} H. Yukawa, {\it Proc. Phys. Math. Soc. Jpn.} {\bf 17}, 48 (1935).
 
 \bibitem{bd13} D. Borka, P. Jovanovi´c, V. Borka Jovanovicetal., {\it JCAP} {\bf 11}, 050 (2013).
 
 \bibitem{af16} A. F. Zakharov, P. Jovanovic, D. Borka et al., {\it JCAP} {\bf 05}, 045 (2016).
 
 \bibitem{af18} A. F. Zakharov, P. Jovanovic, D. Borka et al., {\it JCAP} {\bf 04}, 050 (2018).
 
 \bibitem{sc20} S. Capozziello, V.B. Jovanovic, D. Borka et al., {\it Phys. Dark Univ.} {\bf 29}, 100573 (2020).
 
 \bibitem{id18} I. DeMartino, R. Lazkoz, M. DeLaurentis, {\it Phys. Rev. D} {\bf 97}, 104067 (2018).
 
 \bibitem{md18} M. DeLaurentis, I. DeMartino, R. Lazkoz, {\it Phys. Rev. D} {\bf 97}, 104068 (2018).
 
 \bibitem{jw15} J. W. Moffat, {\it Eur. Phys. J. C} {\bf 75}, 175 (2015).
 
 \bibitem{aj22} A. Jawad, M. B. Amin. Sulehri, S. Rani, {\it Eur. Phys. J. Plus} {\bf 137}, 1274 (2022).

 \bibitem{rg21} R. Garattini, {\it Eur. Phys. J. C} {\bf 81}, 824 (2021).
 
 \bibitem{mb15} M. Bordag, G. L. Klimchitskaya, U. Mohideen, and V. M. Mostepanenko, {\tt Advances in the Casimir Effect}, first ed., Oxford Science Publications, Oxford, (2015).
 
\bibitem{rg20} R.Garattini, {\it Eur. Phys. J. C} {\bf 80},1172 (2020).

\bibitem{mv03} M. Visser, S. Kar, N. Dadhich, {\it Phys. Rev. Lett.} {\bf 90}, 201102 (2003).

\bibitem{dl07} D. Lynden-Bell, J. Katz, and J. Bicak,  {\it Phys. Rev. D} {\bf 75}, 024040 (2007).

\bibitem{kk09} K. K. Nandi, Y. Z. Zhang, R. G. Cai, and A. Panchenko, {\it Phys. Rev. D} {\bf  79}, 024011 (2009).

\bibitem{cw73} C. W. Misner, K. S. Thorne, and J. A. Wheeler, {\tt Gravitation}, San Francisco, (1973).

\bibitem{jm72} J. M. Bardeen, W. H. Press, and S. A. Teukolsky, {\it Astrophys. J.} {\bf 178}, 347 (1972).

\bibitem{se04} S. E. Vazquez and E. P. Esteban, {\it Nuovo Cimento B Serie} {\bf 119}, 489
(2004).

\bibitem{kh09} K. Hioki and K. Maeda, {\it Phys. Rev. D} {\bf 80}, 024042 (2009).

\bibitem{pg13} P. G. Nedkova, V. Tinchev, and S. S. Yazadjiev, {\it Phys. Rev. D} {\bf 88},
124019 (2013).

\bibitem{ks98} K. S. Virbhadra, D. Narasimha, and S. M. Chitre, {\it Astron. Astro. Phys.} {\bf 337}, 1 (1998).

\bibitem{cm01} C. M. Claudel, K. S. Virbhadra, and G. F. R. Ellis, {\it J. Math. Phys.} {\bf 42}, 818 (2001).

\bibitem {vb02} V. Bozza, {\it Phys. Rev. D} {\bf 66}, 103001 (2002).

\bibitem {vb01} V. Bozza et al., {\it Gen. Relativ. Gravit.} {\bf 33}, 1535 (2001).

\bibitem{EA24} E. Cartan, {\it Ann. Ec. Norm. Suppl.} {\bf 41}, 1 (1924).

\bibitem{EA25} E. Cartan, {\it Ann. Ec. Norm. Suppl}. {\bf 42}, 17 (1925).

\bibitem{FW07} F.W. Hehl and Yu. N. Obukhov, {\it Ann. Fond. Louis Broglie} {\bf 32}, 157 (2007).

\bibitem{vn85} V. N. Ponomarev, A.O. Barvinsky, and Yu.N. Obukhov, {\tt Geometrodynamics Methods and Gauge Approach in the Theory of Gravity} (Energoatomizdat, Moscow, 1985, in Russian).

\bibitem{av06} A.V. Minkevich and A.S. Garkun, {\it Class. Quantum Grav.} {\bf 23}, 4237 (2006).

\bibitem{MR17} M. R. Mehdizadeh, A. H. Ziaie, {\it Phys. Rev. D}, {\bf 95 }, 064049 (2017).

\bibitem{MR17A} M. R. Mehdizadeh, A. H. Ziaie, {\it Phys. Rev. D}, {\bf 96 }, 124017 (2017).

\bibitem{ns24} N. Sarkar, S. Sarkar, F. Rahaman, P. Balo, {\it  Eur. Phys. J. Plus}, {\bf 139}, 1-16 (2024).

\bibitem{MR19} M. R. Mehdizadeh, A. H. Ziaie, {\it Phys. Rev. D}, {\bf 99 }, 064033 (2019).

\bibitem{KA16} K. A. Bronnikov and A. M. Galiakhmetov, {\it Phys. Rev. D}, {\bf 94 }, 124006 (2016).



\end{thebibliography}
\end{document}